\documentclass[prd,superscriptaddress,amsfonts,amssymb,amsmath,showpacs,twocolumn,nofootinbib]{revtex4-2}
\usepackage{bm}
\usepackage{amsfonts}
\usepackage{latexsym}
\usepackage{graphicx}
\usepackage{amsmath}
\usepackage{palatino}
\usepackage{xcolor} 
\usepackage{mathpazo}
\usepackage{xcolor}
\usepackage{rotating}
\usepackage{adjustbox}
\usepackage{tensor}
\usepackage{textcomp}
\usepackage{float}
\usepackage{booktabs}
\usepackage{dcolumn}
\usepackage[titletoc]{appendix}
\usepackage{booktabs}
\usepackage{multirow}
\usepackage{hyperref}
\hypersetup{colorlinks,citecolor=blue}
\usepackage{amsmath}
\usepackage{xcolor}
\usepackage{orcidlink}
\usepackage{epsfig}
\usepackage{caption}
\usepackage{subcaption}
\usepackage{commath}
\hypersetup{colorlinks,citecolor=blue}
\hypersetup{colorlinks=true,linkcolor=magenta,filecolor=magenta,    urlcolor=blue}

\hypersetup{
    colorlinks = true,
    linkcolor = blue,
    anchorcolor = blue,
    citecolor = blue,
    filecolor = blue,
    urlcolor = blue,
    breaklinks=true
   }

\def\be{\begin{equation}}
\def\ee{\end{equation}}
\def\bea{\begin{eqnarray}}
\def\eea{\end{eqnarray}}

\def\be{\begin{equation}}
\def\ee{\end{equation}}
\def\bea{\begin{eqnarray}}
\def\eea{\end{eqnarray}}

\renewcommand{\bar}{\overline}

\begin{document}

\title{Schr\"{o}dinger-type $f(Q,T)$ gravity-nonmetricity driven cosmological evolution from inflation to the late Universe}

\author{Lei Ming}
\email{minglei@mail.sysu.edu.cn}
\affiliation {Key Laboratory of Atomic and Subatomic Structure and Quantum Control (Ministry of Education), Guangdong Basic Research Center of Excellence for Structure and Fundamental Interactions of Matter, School of Physics, South China Normal University, Guangzhou 510006, China} 

\affiliation {Guangdong Provincial Key Laboratory of Quantum Engineering and Quantum Materials, Guangdong-Hong Kong Joint Laboratory of Quantum Matter, South China Normal University, Guangzhou 510006, China}

\author{Himanshu Chaudhary}
\email{himanshuch1729@gmail.com}
\affiliation{Department of Physics, Babe\c s-Bolyai University, Kog\u alniceanu
Street, Cluj Napoca, 400084, Romania}

\author{Shi-Dong Liang}
\email{stslsd@mail.sysu.edu.cn}
\affiliation{School of Physics,
Sun Yat-Sen University, Guangzhou 510275, People’s Republic of China,}

\author{Hong-Hao Zhang}
\email{zhh98@mail.sysu.edu.cn}
\affiliation{School of Physics,
Sun Yat-Sen University, Guangzhou 510275, People’s Republic of China,}

\author{Tiberiu Harko}
\email{tiberiu.harko@aira.astro.ro}
\affiliation{Department of Physics, Babe\c s-Bolyai University, Kog\u alniceanu
Street, Cluj Napoca, 400084, Romania}
\affiliation{Astronomical Observatory,
19 Cire\c silor Street, Cluj-Napoca 400487, Romania}

\begin{abstract}
We consider an $f(Q, T)$ gravity theory with a Schr\"{o}dinger type vectorial non-metricity. In the presence of such a non-metricity, the length of vectors is preserved under autoparallel transport. We obtain the field equations assuming a vanishing total scalar curvature, implemented by a Lagrange multiplier, and investigate their cosmological implications. To do this, we derive the generalized Friedmann equations which now have terms involving the non-metricity and the Lagrange multiplier. Then, we consider two distinct cosmological applications of the model. First of all, by adopting distinct forms of these two basic variables and investigate the possibility of the existence of warm inflationary scenarios within the framework of these models. In particular,  we consider the case that the non-metricity is described by a constant vector, and we show that with this assumption we recover standard general relativity.  The scenario in which the Lagrange multiplier is a constant is also investigated, and we show that radiation can be created during the very early phases of expansion. The amount of radiation peaks at a certain time after which, there is a transition from an accelerating inflationary phase to a decelerating one. Moreover, we perform a detailed comparison of the predictions of the considered Schr\"{o}dinger type cosmology with a set of observational data for the Hubble function, including Cosmic Chronometers, Type Ia Supernovae, and Baryon Acoustic Oscillations, using a Markov Chain Monte Carlo (MCMC) analysis, by adopting a simple linear form for the Lagrange density. The model predictions are also compared with the results of the $\Lambda$CDM standard paradigm. Our results indicate that the Schr\"{o}dinger  $f(Q,T)$ type theory can give a good description of the observational data for both the very early and the late Universe.
\end{abstract}

\date{\today }
\maketitle
\tableofcontents

\affiliation{Department of Physics, Babes-Bolyai University, Kogalniceanu
Street, Cluj Napoca, 400084, Romania}

\affiliation{ School of Physics,
Sun Yat-Sen University, Guangzhou 510275, People’s Republic of China,}

\affiliation{ School of Physics,
Sun Yat-Sen University, Guangzhou 510275, People’s Republic of China,}

\affiliation{ School of Physics,
Sun Yat-Sen University, Guangzhou 510275, People’s Republic of China,}

\affiliation{Department of Physics, Babes-Bolyai University, Kogalniceanu
Street, Cluj Napoca, 400084, Romania}
\affiliation{Astronomical Observatory,
19 Ciresilor Street, Cluj-Napoca 400487, Romania}

{\ \hypersetup{linkcolor=blue} }

\section{Introduction}\hypersetup{citecolor=blue}

The discovery of the recent acceleration of the Universe \cite{acc1,acc2,acc3,acc4,acc5} has dramatically changed the landscape of cosmology and of the gravitational theories. The findings of the Planck satellite \cite{Planck} have also revealed previously unexpected properties and composition of the Universe. In order to explain the observational data, the $\Lambda$CDM cosmological model \cite{Peebles1} was proposed, which was quite successful in describing cosmological observations. Based on the interpretation of the observations, the $\Lambda$CDM model assumes a spatially flat Universe, composed by approximately 5\% ordinary matter, 25\% cold dark matter (CDM), and 70\% dark energy (DE), described  by a cosmological constant $\Lambda$. However, recently several important deviations between the predictions of the $\Lambda$CDM model and observations did arise. Besides the standard and old problems of the cosmological constant \cite{Weinberg} and of dark matter, several new disparities did appear when comparing the $\lambda$CDM model with the most recent cosmological data. These new problems include the Hubble and $S_8$ tensions, as well as disagreements related to the validity of the cosmological principle \cite{Riess, Val1, Val2, Aluri, Peebles, Riess1, Asgari}. All these observational results and the associated cosmological tensions points towards the possibility that the $\Lambda$CDM standard model may be just an approximation of a more general cosmological model, based on a gravitational theory that extends and modifies general relativity (GR) \cite{Ein, Hilb}, the dominant present day theory of gravity. 

On the other hand, the evolution of the very early Universe is assumed to be described, within the theory of general relativity,  by the inflationary theory \cite{Guth, Baumann, Cline}, which assumes a de Sitter type initial expansion, followed by a reheating period, during which matter, mostly in the form of radiation, was created from the decay of the scalar field that triggered the exponentially expanding phase.

To explain the recent cosmological dynamics several extensions and modifications of GR have been proposed in the literature. One such very promising approach is using non-Riemannian geometry to formulate novel gravitational theories. From a purely mathematical perspective, a four-dimensional curved manifold can be characterized by three independent quantities: curvature \cite{Riemann}, non-metricity \cite{Weyl1,Weyl2,Weyl3,Weyl4, Weyl5}, and torsion \cite{C1,C2,C3, C4,Hehl, TEGR1,TEGR2,  Csillag2}. 

Each of these basic geometric quantities can be used independently (or in some combination) to describe the gravitational interaction. Using only non-metricity, an equivalent formulation of GR  was considered in \cite{Q1}. The generalization of this approach resulted in the development of the symmetric teleparallel gravity theory, which was further extended into the $f(Q)$ gravity theory \cite{Q2}. In this approach to gravity the basic geometric quantity is the non-metricity scalar $Q$ (for a recent review of $f(Q)$ type theories see \cite{Q3}). The coupling of matter with non-metricity was introduced in \cite{Q4}, and this approach was later extended to the  $f(Q, T )$ gravity theory \cite{Q5,Q6}.  

 The action of the $f(Q,T)$ gravity theory is given in its most general form by \cite{Q5}  
 \be
 S=\int{f(Q,T)\sqrt{-g}d^4x}+S_m, 
 \ee
 where $T$ is the trace of the matter energy-momentum tensor, and $S_m=\int{L_m\sqrt{-g}d^4x}$ is the matter action, with $L_m$ denoting the matter Lagrange density. A particular formulation of the $f(Q,T)$ gravity theory, as well as of the $f(Q)$ theory, is the Weyl-type $f(Q,T)$ gravity \cite{Q6}, where it is explicitly assumed that the non-metricity is of Weyl type \cite{Weyl1,Weyl2}. The action of the  Weyl-type $f(Q,T)$ gravity is formulated as \cite{Q6}
\bea
 \hspace{-0.5cm}S=\int\sqrt{-g}d^4x\Bigg[&&\kappa ^2f(Q,T)-\frac{W_{\mu\nu}W^{\mu \nu}}{4}-\frac{m^2 w_\mu w^\mu}{2}\nonumber\\
 \hspace{-0.5cm}&&+\lambda \left(R+6\nabla _{\mu} w^{\mu}-6w_{\mu} w^{\mu}\right)+L_m\Bigg],
 \eea 
 where $w_\mu$ denotes the Weyl vector field, $m$ is a constant (the mass of the Weyl vector field), and $W_{\nu \mu}=\nabla _\mu w_\nu-\nabla _\nu w_\mu$ is the strength of the Weyl vector field.  The important condition of the vanishing of the total curvature of the Weyl type manifold is introduced with the use of the Lagrange multiplier $\lambda$. The considered coupling between geometry and matter is similar to the one previously considered in the $f(R,T)$ gravity theory \cite{fRT}, and its extensions \cite{book}. 
 
 The general astrophysical, physical and cosmological implications of the $f(Q,T)$ and Weyl type $f(Q,T)$ theories were extensively analyzed in the recent literature \cite{P1, P2,P3,P4,P5,P6,P7,P8,P9,P10,P11,P12,P13,P14,P15,P16,P17,P18,P19,P20,P21,P22,P23,P24,P25,P26,P27,P28,P29,P30,P31,P32,P33,P34,P35, P36,P37,P38,P39,P40,P41,P42, P43,P44,P45}. 

 Initially, the introduction of the Weyl-type vectorial non-metricity \cite{Weyl1,Weyl2} was motivated by the hope of finding a theory, which unifies electromagnetism and gravity. However, Weyl's attempt was heavily criticized by Einstein \cite{Pala}, which has led physicists to neglect Weyl geometry for more than a century \cite{Weyl3}. However, recently it was found that Weyl geometry is useful in a large range of physical applications, ranging from  condensed matter theory \cite{Palumbo2024} to elementary particle physics \cite{Ghil1,Ghil2}.

In the present work, we consider an $f(Q,T)$ type theory with the underlying geometry equipped with a scalar flat Schr\"odinger connection. This connection allows for the lengths of vectors to be preserved under autoparallel transport \cite{Sch1,Sch2}.  In light of such physically appealing properties, there has been a resurgence of interest in Schr\"{o}dinger geometry recently. It was studied both in the metric-affine \cite{Rav,Csillag2024} and metric \cite{Ming}  formalisms. In particular, in \cite{Ming}, a gravitational theory with action 
\begin{align}\label{action}
S=\frac{1}{16\pi}\int d^4x& \sqrt{-g}\bigg(R+\frac{5}{24}Q_\rho Q^\rho+\frac{1}{6}\tilde{Q}_\rho\tilde{Q}^\rho+2T_\rho Q^\rho\nonumber\\
&+\zeta^{\rho\sigma}_{~~\alpha}T^\alpha_{~\rho\sigma}\bigg)+\int d^4x\sqrt{-g}L_m,
\end{align}
was considered, where $R= g^{\mu\nu}R_{\mu\nu}(\Gamma)$, $T_\rho= T^\sigma_{~\rho\sigma}$ is the torsion, and $\zeta^{\rho\sigma}_{~~\alpha}$ is a Lagrange multiplier. In the metric version  the theory contributes some non-metricity dependent extra terms in the gravitational Einstein equations, which can be interpreted as representing a geometric type dark energy. After obtaining the generalized Friedmann equations,  the cosmological implications of the theory were analyzed in detail, by considering two distinct models, corresponding to a dark energy satisfying a linear equation of state, and to conserved matter energy, respectively. 

\begin{figure*}[htbp]
\includegraphics[width=0.4\linewidth]{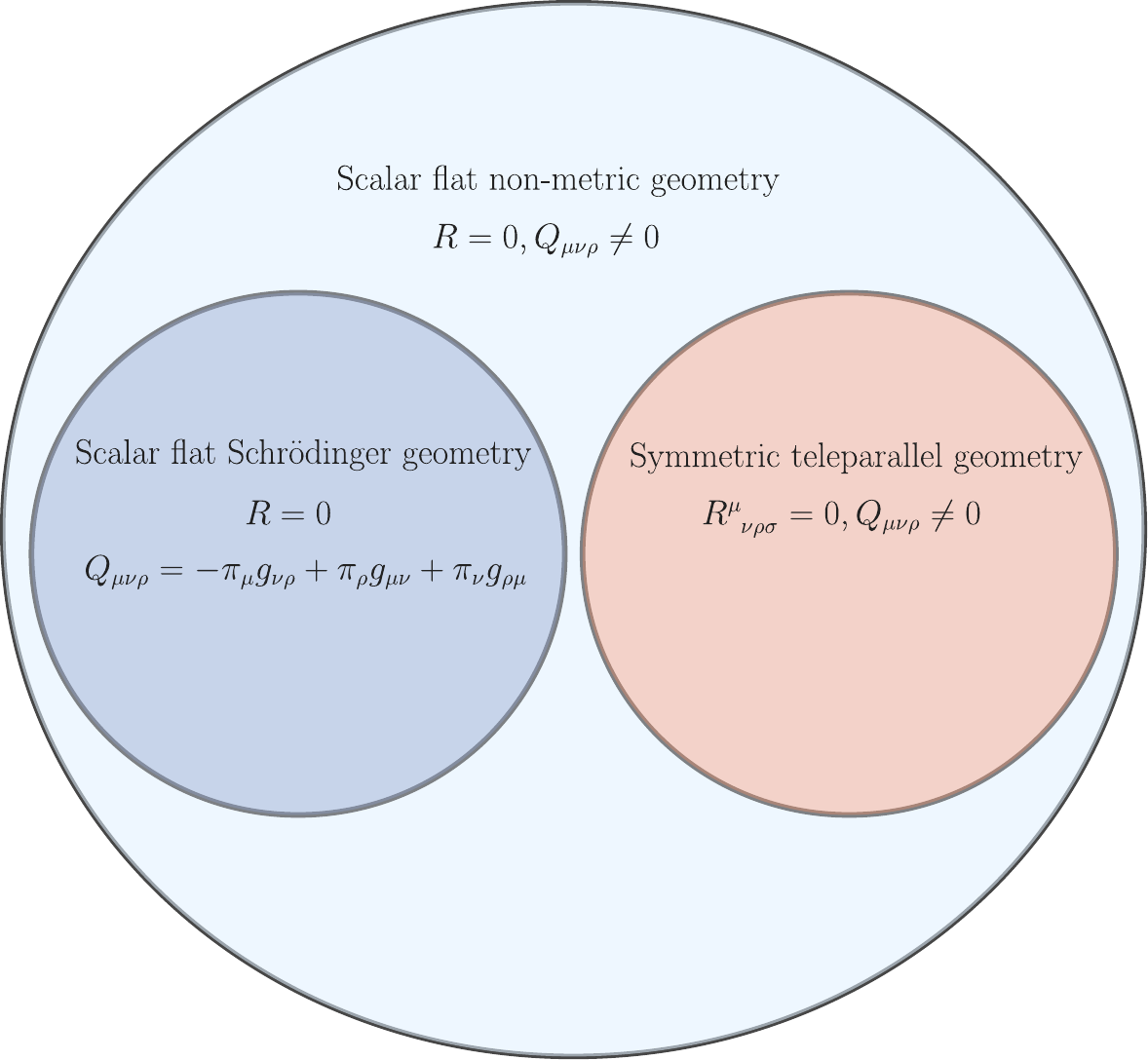}
\caption{Illustration of several flat geometries with nonmetricity.}
\label{Schrodingerdrawing}
\end{figure*}

In our present approach to the physical applications of the Schr\"{o}dinger geometry we consider an action analogous to the one used in \cite{Q6}, but in terms of a vectorial Schr\"odinger non-metricity (as opposed to a Weyl-type). To obtain a closed theory in which non-metricity is dynamical, we assume a flat geometry and impose the condition of vanishing total scalar curvature using a Lagrange multiplier. By "scalar flat," we mean that the total Ricci scalar is zero. It is important to note that vanishing scalar curvature does not imply a vanishing Riemann tensor, although the converse is true. Typically $f(Q)$ theories are studied within the symmetric teleparallel framework, where flatness is imposed at the level of the full Riemann tensor $\tensor{R}{^\mu _\nu _\rho _\sigma}$, not just the Ricci scalar $R$. However, this approach is generally not viable for the Schr\"{o}dinger connection, as a coincident gauge for such connections generally does not exist. The geometric assumptions underlying our construction are illustrated in Fig.~\ref{Schrodingerdrawing}.

As an application of the new gravitational theory developed here, we consider the dynamics of a spatially flat Friedmann-Lemaitre-Robertson-Walker (FLRW) Universe. Using the metric field equations, we first obtain the generalized Friedmann equations, which contain extra terms that depend on the Schr\"{o}dinger vector and the Lagrange multiplier. These extra terms can be interpreted as generating an effective dark energy, with the effective energy density and pressure determined by non-metricity and the Lagrange multiplier. Then, we turn our attention to the paradigm of cosmic inflation \cite{Guth, Baumann, Cline}. One of the important features of the present $f(Q,T)$ theory is that the matter energy-momentum tensor is not conserved. This property can be interpreted physically as describing matter creation from the geometry. Based on this interpretation, we consider a warm inflationary scenario within the framework our Schr\"{o}dinger $f(Q,T)$ gravity. 

Warm inflation is an alternative scenario to the standard inflation and reheating models, in which during the inflationary epoch the inflaton interacts dynamically with other fields through thermal dissipation, allowing for energy transfer between the inflation and other fields \cite{Fang, Be1,Be2, Be3}. 
Due to this important property, a sudden reheating mechanism, required in the cold inflationary models, is not necessary for warm inflation. Moreover, the model also allows for a  smooth transition of the inflaton field to a state where its effect on the expansion of the Universe is negligible. For recent investigations in the field of warm inflation, see \cite{W1,W2,W3,W4,W5,W6,W7,W8,W9,W10,W11,W12,W13,W14,W15,W16,W17,W18,W19,101,102}.

Inspired by the warm inflationary scenario we investigate an alternative proposal for the radiation creation in the early Universe, in which the photon fluid is created from the geometry, and in particular from the Schr\"{o}dinger vector. The source term in the energy-momentum balance equation triggers the creation of radiation, which is produced due to the presence of Schr\"{o}dinger vector. Two distinct theoretical models for radiation creation are considered. We find that in all of the models, the Universe begins in a state without radiation. The radiation is produced during the inflationary epoch. The energy density of the  radiation reaches a maximum, and decreases afterwards, when the expansion rate becomes greater than the particle production rate.

As a second application of the proposed theoretical model we consider the late Universe tests of the Schr\"{o}dinger type $f(Q,T)$ theory. For simplicity we assume a simple functional form for $f(Q,T)$, which we take as the sum of the nonmetricity scalar $Q$ and the trace of the matter energy-momentum tensor $T$. Then the theoretical predictions of the Schr\"{o}dinger type $f(Q,T)$ model are compared against several observational datasets, including Cosmic Chronometers, Type Ia Supernovae, and Baryon Acoustic Oscillations, using a Markov Chain Monte Carlo (MCMC) analysis, which allows the determination of the two free parameters of the model. A comparison with the predictions of standard  $\Lambda$ CDM model is also performed. Our results indicate that the simple Schr\"{o}dinger type $f(Q,T)$ model gives a relatively good description of the observational data, and thus it can be considered as a successful, and simple, alternative to standard cosmological models. 

The present paper is organized as follows. We develop the basic geometric concepts and introduce the action of Schr\"{o}dinger type $f(Q,T)$ gravity in Section~\ref{sect1}. The evolution of a FLRW-type flat Universe is considered in Section~\ref{sect2}, where the generalized Friedmann equations, including the evolution equations for Schrödinger\"{o}dinger vector and the Lagrange multiplier are derived. Warm inflationary type models describing the creation of radiation from geometry are considered in Section~\ref{sect3}. The late time evolution of the Universe in the framework of a simple Schr\"{o}dinger type $f(Q,T)$ model is considered in Section~\ref{sect4}. We discuss and conclude our results in Section~\ref{sect5}. Some basic geometric relations are presented in Appendix~\ref{appendixA}. The technical details of the derivation of the field equations and of the Friedmann equations are given in Appendices ~\ref{appendixB} and ~\ref{appendixC}, respectively.                 

\section{Schr\"{o}dinger-type $f(Q,T)$ gravity}\label{sect1}

In this section, we present a generalization of symmetric teleparallel gravity by using the Schr\"{o}dinger geometry, which admits fixed length vectors. We first review the basics of the Schr\"{o}dinger geometry, then turn present the field equations of the proposed theory and obtain the divergence of the energy-momentum tensor, a quantity that is not conserved in the present theory.

\subsection{Schr\"{o}dinger geometry}

Inspired by \cite{Jimenez2016}, in which a generalized Weyl geometry was introduced, one could study linear vectorial connections of the form
\begin{equation}\label{vectorialdistortion}
    \tensor{\Gamma}{^\mu_\nu_\rho}=\overset{\circ}{\Gamma}\tensor{}{^\mu _\nu _\rho} -b_1 \pi^{\mu} g_{\nu \rho}+\frac{b_2}{2}\delta^{\mu}_{\nu} \pi_\rho +\frac{b_2}{2} \delta^{\mu}_{\rho} \pi_{\nu},
\end{equation}
where $\overset{\circ}{\Gamma}\tensor{}{^\mu _\nu _\rho}$ denotes the Levi-Civita connection, and $\pi$ is a one-form. Working in the conventions of \cite{Csillag2024b}, a calculation presented in Appendix \ref{appendixA} reveals that this is a connection with non-metricity given by
\begin{equation}\label{vectorialnon-metricity}
    Q_{\mu \nu \rho}=b_2 \pi_{\mu} g_{\nu \rho} + \frac{b_2 -2 b_1}{2} \left( \pi_{\rho} g_{\mu \nu} + \pi_{\nu} g_{\rho \mu} \right).
\end{equation}

It was proven in \cite{Ming} that a sufficient condition for a nonmetric geometry to admit fixed length vectors is $Q_{(\mu \nu \rho)}=0$. A calculation detailed in Appendix \ref{appendixA} shows that this condition  imposes the constraint $b_1=b_2$ on the coefficients of the linear vectorial distortion.

For simplicity, we fix $b_1=b_2=-1$ and proceed with the Yano-Schr\"{o}dinger geometry
\begin{equation}  \label{thisconnection}
\tensor{\Gamma}{^\mu _\nu _\rho}=\overset{\circ}{\Gamma} \tensor{}{^\mu _\nu _\rho}+ \pi^{\mu}
g_{\nu \rho} - \frac{1}{2} \pi_{\rho} \delta^{\mu}_{\nu} - \frac{1}{2}
\pi_{\nu} \delta^{\mu}_{\rho}.
\end{equation}
Hence, we consider a geometry characterized by a distortion tensor
\begin{equation}\label{distortion}
    \tensor{N}{^\mu _\nu _\rho}=\tensor{\Gamma}{^\mu _\nu _\rho} - \overset{\circ}{\Gamma}\tensor{}{^\mu _\nu _\rho}=\pi^{\mu}
g_{\nu \rho} - \frac{1}{2} \pi_{\rho} \delta^{\mu}_{\nu} - \frac{1}{2}
\pi_{\nu} \delta^{\mu}_{\rho}.
\end{equation}

The Ricci tensor for a connection with distortion is generally given by
\begin{multline}\label{Riccitensorgeneral}
    R_{\mu \nu}=\overset{\circ}{R}_{\mu \nu}+ \overset{\circ}{\nabla}_{\alpha}\tensor{N}{^\alpha _\nu _\mu} - \overset{\circ}{\nabla}_{\nu} \tensor{N}{^\alpha _\alpha _\mu}\\
    + \tensor{N}{^\alpha _\alpha _\rho} \tensor{N}{^\rho _\nu _\mu} - \tensor{N}{^\alpha _\nu _\rho} \tensor{N}{^\rho _\alpha _\mu}.
\end{multline}

Substituting the expression for the distortion tensor \eqref{distortion} into \eqref{Riccitensorgeneral}, we obtain the modified Ricci tensor \cite{Csillag2024}
\begin{eqnarray}  \label{Riccitensor}
\begin{aligned} R_{\mu \nu}&=&\overset{\circ}{R}_{\mu \nu}+ g_{\mu \nu}
\overset{\circ}{\nabla}_{\alpha} \pi^{\alpha} - \frac{1}{2}
\overset{\circ}{\nabla}_{\mu} \pi_{\nu} + \overset{\circ}{\nabla}_{\nu}
\pi_{\mu} \nonumber\\ &&- \frac{1}{2} g_{\mu \nu} \pi^\alpha \pi_\alpha
-\frac{1}{4} \pi_\mu \pi_\nu. \end{aligned}
\end{eqnarray}

The corresponding Ricci scalar takes the form
\begin{equation}  \label{Ricciscalar}
R=\overset{\circ}{R}+\frac{9}{2} \overset{\circ}{\nabla}_{\mu} \pi^\mu -
\frac{9}{4} \pi_\mu \pi^\mu.
\end{equation}

The building block of our theory is the non-metricity scalar, defined as
\begin{equation}  \label{non-metricityscalar}
Q=-Q_{\alpha \mu \nu}P^{\alpha \mu \nu},
\end{equation}
where the superpotential $P^{\alpha \mu \nu}$ is expressed as
\begin{multline}  \label{superpotential}
P^{\alpha \mu \nu}=-\frac{1}{4} Q^{\alpha \mu \nu} + \frac{1}{2} Q^{(\mu
\nu)\alpha}+\frac{1}{4} \left (Q^\alpha - \widetilde{Q}^\alpha \right)
g^{\mu \nu} \\- \frac{1}{4} g^{\alpha( \mu} Q^{\nu)},
\end{multline}
with the non-metricity traces being
\begin{equation}  \label{non-metricityvector}
Q_{\alpha}=g^{\sigma \lambda} Q_{\alpha \sigma \lambda}, \; \; \widetilde{Q}%
_{\alpha}=g^{\sigma \lambda} Q_{\sigma \alpha \lambda}.
\end{equation}

For the Schr\"{o}dinger connection introduced in \eqref{thisconnection}, a direct calculation shown in Appendix \ref{appendixA} yields
\begin{equation}\label{non-metricityscalarschrexpression}
Q=-\frac{9}{4} \pi_{\alpha} \pi^{\alpha}.
\end{equation}

Therefore, the non-metricity scalar, and consequently the entire non-metricity contribution to the gravitational action, is fully determined by the Schrödinger vector $\pi^{\mu}$. This highlights the central role played by the vector field $\pi^{\mu}$ in encoding the geometric degrees of freedom of the theory.

\subsection{Variational principle and field equations}

With the geometric foundations established, we formulate the foundations of Schr\"{o}dinger-type $f(Q,T)$ gravity. The action of this theory is given by
\begin{equation}\label{thisaction}
\begin{aligned} S=\int \mathrm{d}^4x \sqrt{-g} \Bigg[ &\kappa^2 f(Q,T) -
\frac{1}{4} \Pi_{\mu \nu} \Pi^{\mu \nu}- \frac{1}{2} m^2 \pi_\mu \pi^\mu \\
&+ \mathcal{L}_{m}+\lambda \left(\overset{\circ}{R} + \frac{9}{2}
\overset{\circ}{\nabla}_{\mu} \pi^{\mu} - \frac{9}{4} \pi_\mu \pi^\mu
\right) \Bigg], \end{aligned}
\end{equation}
where we have defined
\begin{equation}
\Pi_{\mu \nu} := \overset{\circ}\nabla_{\nu} \pi_{\mu} - \overset{\circ}\nabla_{\mu} \pi_{\nu}.
\end{equation}

The present theory includes a non-minimal coupling between the Schrödinger vector field $\pi^\mu$  and matter through the $f(Q,T)$ term, which couples $\pi^{\mu}$ to the trace of the energy-momentum tensor of matter. The non-metricity vector field becomes dynamical through the addition of a Maxwell-type kinetic term with $\Pi_{\mu \nu}=\overset{\circ}{\nabla}_{\nu} \pi_{\mu} - \overset{\circ}{\nabla}_{\mu} \pi_{\nu}$, and gains mass through a Proca-like term $\pi_{\mu} \pi^{\mu}$. Additionally, the Lagrange multiplier $\lambda$ imposes the scalar flat constraint
\begin{equation}
\overset{\circ}{R} + \frac{9}{2} \overset{\circ}{\nabla}_{\mu} \pi^{\mu} -
\frac{9}{4} \pi_\mu \pi^\mu=0 \iff R=0,
\end{equation}
where $R$ denotes the Ricci scalar of the Schr\"{o}dinger connection.

The field equations obtained from the action \eqref{thisaction} take the form 
\begin{widetext}
\begin{multline}\label{fieldeqn1}
    \frac{1}{2} T_{\mu \nu} + \frac{1}{2} \Pi_{\mu \beta} \tensor{\Pi}{_\nu ^\beta} - \frac{1}{8} \Pi_{\rho \lambda} \Pi^{\rho \lambda} g_{\mu \nu} +\frac{1}{2} m^2 \pi_\mu \pi_\nu {-} \frac{1}{4} m^2 g_{\mu \nu} \pi_{\rho} \pi^{\rho}-\kappa^2 f_T \left( -T_{\mu \nu} + g_{\mu \nu} L_{m} \right)\\
    =
     - \frac{9}{4} \kappa^2 f_Q \pi_\mu \pi_\nu- \frac{1}{2} \kappa^2 g_{\mu \nu} f + \lambda \left( \overset{\circ}{R}_{\mu \nu} - \frac{9}{4} \pi_{\mu} \pi_{\nu} -\frac{1}{2} g_{\mu \nu} \overset{\circ}{R} +\frac{9}{8} g_{\mu \nu} \pi_{\rho} \pi^{\rho} \right) + \frac{9}{4} g_{\mu \nu} \pi^{\rho} \overset{\circ}{\nabla}_{\rho}  \lambda\\
     - \frac{9}{4} \pi_{\mu} \overset{\circ}{\nabla}_{\nu} \lambda -\frac{9}{4} \pi_{\nu} \overset{\circ}{\nabla}_{\mu} \lambda + g_{\mu \nu} \overset{\circ}{\Box} \lambda - \overset{\circ}{\nabla}_{\mu} \overset{\circ}{\nabla}_{\nu} \lambda,
\end{multline}
\begin{equation}\label{vectorfieldeom}
     {\overset{\circ}{\Box} \pi_{\mu} -
     \overset{\circ}{\nabla}_{\nu} \overset{\circ}{\nabla}_{\mu} \pi^{\nu}}  -\left(m^2+\frac{9}{2}\kappa^2 f_Q {+}\frac{9}{2} \lambda \right) \pi_{\mu}=\frac{9}{2} \overset{\circ}{\nabla}_{\mu} \lambda.
\end{equation}
\end{widetext}

\subsection{Dynamical evolution in a flat FLRW metric}\label{sect2}

As a first step in investigating cosmological implications, we evaluate the field equations in a spatially flat FLRW metric, given by
\begin{equation}
\mathrm{d}s^{2}=-\mathrm{d}t^{2}+a^{2}(t)\left(dx^2+dy^2+dz^2 \right),
\end{equation}
where $a(t)$ is the scale factor. Due to homogeneity and isotropy, we take the Lagrange multiplier to be a function of time only, $\lambda =\lambda (t)$. Furthermore, the  Schr\"odinger  vector is assumed to have the form
\begin{equation}
\pi _{\mu }=\left[ \pi (t),0,0,0\right] .
\end{equation}

Assuming the matter content of the universe is described by a perfect fluid, the nonvanishing components of the energy-momentum tensor take the form (here and in the following $i=1,2,3$)
\begin{equation}
T_{00}=\rho ,\quad T_{ii}=pa^{2},
\end{equation}%
where $\rho $ and $p$ are the energy density and the pressure, respectively.

Based on these assumptions, the following relations are immediate
\begin{equation}
\Pi_{\mu\nu}=0, \quad \pi_\mu\pi^\mu=-\pi^2, \quad \mathcal{L}_m=p.
\end{equation}

After lengthy algebra, highlighted in Appendix \ref{appendixC}, we obtain the following set of differential equations
\begin{eqnarray}
3H^{2}=\frac{\rho }{2\lambda }+\frac{1}{\lambda }\Bigg[ &&\kappa
^{2}f_{T}(\rho +p)-\frac{\kappa ^{2}f}{2}+\frac{9}{4}\kappa ^{2}f_{Q}{\pi
^{2}}+\frac{1}{4}m^{2}\pi ^{2}  \notag \\
&&+\frac{9}{8}{\lambda }\pi ^{2}+\frac{9}{4}\dot{\lambda}\pi -3\dot{\lambda}H%
\Bigg],  \label{Fr1}
\end{eqnarray}%
\begin{eqnarray}
2\dot{H}+3H^{2} &=&-\frac{p}{2\lambda }+\frac{1}{\lambda }\Bigg(-\frac{%
\kappa ^{2}f}{2}-\frac{1}{4}m^{2}\pi ^{2}-\frac{9}{8}\lambda \pi ^{2}  \notag
\\
&&-\frac{5}{4}\dot{\lambda}\pi -2\dot{\lambda}H+\frac{2}{9}m^{2}\dot{\pi}
\notag \\
&&+\kappa ^{2}\dot{f}_{Q}\pi +\kappa ^{2}f_{Q}\dot{\pi}+\lambda \dot{\pi}%
\Bigg),  \label{Fr2}
\end{eqnarray}%
\begin{equation}
\dot{\lambda}=\left( -\frac{2}{9}m^{2}-\kappa ^{2}f_{Q}-\lambda \right) \pi ,
\label{Fr3}
\end{equation}%
\begin{equation}
\dot{\pi}=\frac{8}{3}H^{2}+\frac{4}{3}\dot{H}-3H\pi +\frac{1}{2}\pi ^{2}.
\label{Fr4}
\end{equation}%
%
%
%
%

Equations (\ref{Fr1}) and (\ref{Fr2}) can be written in terms an effective form as
\begin{equation}
3H^{2}=\frac{\rho }{2\lambda }+\frac{\rho _{eff}}{\lambda },  \label{Fr5}
\end{equation}%
and
\begin{equation}
2\dot{H}+3H^{2}=-\frac{p}{2\lambda }-\frac{p_{eff}}{\lambda },  \label{Fr6}
\end{equation}%
where $\rho_{eff}$ and $p_{eff}$ represent the effective energy density and effective pressure of the dark energy component, defined by
\begin{eqnarray}
\rho _{eff}&=&\kappa ^{2}f_{T}(\rho +p)-\frac{\kappa ^{2}f}{2}+\frac{9}{4}%
\kappa ^{2}f_{Q}{\pi ^{2}}+\frac{1}{4}m^{2}\pi ^{2}  \notag \\
&&+\frac{9}{8}{\lambda }\pi ^{2}+\frac{9}{4}\dot{\lambda}\pi -3\dot{\lambda}%
H,  \label{Fr7}
\end{eqnarray}
and
\begin{eqnarray}
p_{eff}&=&\frac{\kappa ^{2}f}{2}+\frac{1}{4}m^{2}\pi ^{2}+\frac{9}{8}\lambda
\pi ^{2}+\frac{5}{4}\dot{\lambda}\pi +2\dot{\lambda}H-\frac{2}{9}m^{2}\dot{%
\pi}  \notag \\
&&-\kappa ^{2}\dot{f}_{Q}\pi -\kappa ^{2}f_{Q}\dot{\pi}-\lambda \dot{\pi},
\label{Fr8}
\end{eqnarray}

From equations (\ref{Fr5}) and (\ref{Fr6}), we obtain the following
dynamical equation for $H$

\begin{equation}
2\dot{H}=-\frac{\rho +p}{2\lambda }-\frac{\rho _{eff}+p_{eff}}{\lambda }.
\label{Fr9}
\end{equation}

By taking the time derivative of Eq. (\ref{Fr5}), we obtain

\begin{equation}
6H\dot{H}=\frac{\dot{\rho}}{2\lambda }-\rho \frac{\dot{\lambda}}{2\lambda
^{2}}+\frac{\dot{\rho}_{eff}}{\lambda }-\rho _{eff}\frac{\dot{\lambda}}{%
\lambda ^{2}}.  \label{Fr10}
\end{equation}

Substituting $2\dot{H}$ with the help of Eq. (\ref{Fr9}), we obtain the
generalized conservation equation in Schrodinger type $f(Q,T)$ gravity as
\begin{eqnarray}
\dot{\rho}+3H\left( \rho +p\right) -\rho \frac{\dot{\lambda}}{\lambda }+2%
\Bigg[ \dot{\rho}_{eff}&+&3H\left( \rho _{eff}+p_{eff}\right)  \notag \\
&&-\rho _{eff}\frac{\dot{\lambda}}{\lambda }\Bigg] =0.  \label{Fr11}
\end{eqnarray}

Equation ~(\ref{Fr11}), which governs the evolution of the matter energy density, must be solved simultaneously with equation ~(\ref{Fr9}), describing the dynamical behavior of the Hubble parameter $H$,
as well as with equations ~(\ref{Fr3}) and (\ref{Fr4}). This system becomes fully determined once the equation of state for the
baryonic matter component is specified.

\subsubsection{Dimensionless representation of the generalized Friedmann equations}

To simplify the mathematical formalism we rewrite the evolution equation
(\ref{Fr1})-(\ref{Fr4}) in a dimensionless form, by introducing the set of
dimensionless variables $\left( \tau ,h,\Pi ,\Lambda ,r,P,F,\bar{Q},\bar{T}\right) $ defined
according to
\begin{eqnarray}\label{47}
\tau  &=&H_{0}t,H=H_{0}h,\pi =H_{0}\Pi ,\lambda =\kappa ^{2}\Lambda
,f=H_{0}^{2}F,\nonumber\\
m^{2}&=&\kappa ^{2}M^{2},
\rho  =\kappa ^{2}H_{0}^{2}r,p=\kappa ^{2}H_{0}^{2}P,Q=H_{0}^{2}\bar{Q},\nonumber\\
T&=&\kappa ^{2}H_{0}^{2}\bar{T}.
\end{eqnarray}%
where $H_{0}$ is a constant reference Hubble scale. Let us note that $\Lambda$ is merely a dimensionless variable, not related to the cosmological constant. Using these definitions, the system of differential equations takes the form
\begin{eqnarray}\label{DFr1}
3h^{2}=\frac{r}{2\Lambda }+\frac{1}{\Lambda }\Bigg[ &&F_{\bar{T}}(r+P)-\frac{%
F}{2}+\frac{9}{4}F_{\bar{Q}}\Pi {^{2}}+\frac{1}{4}M^{2}\Pi ^{2}  \label{DFr1}
\nonumber\\
&&+\frac{9}{8}{\Lambda }\Pi ^{2}+\frac{9}{4}\Pi\frac{d\Lambda }{d\tau } -3h%
\frac{d\Lambda }{d\tau }\Bigg],
\end{eqnarray}

\begin{eqnarray}\label{DFr2}
2\frac{dh}{d\tau }+3h^{2} &=&-\frac{P}{2\Lambda }+\frac{1}{\Lambda }\Bigg(-%
\frac{F}{2}-\frac{1}{4}M^{2}\Pi ^{2}-\frac{9}{8}\Lambda \Pi ^{2}  \notag \\
&&-\frac{5}{4}\frac{d\Lambda }{d\tau }\Pi -2\frac{d\Lambda }{d\tau }h+\frac{2%
}{9}M^{2}\frac{d\Pi }{d\tau }  \notag \\
&&+\frac{dF_{\bar{Q}}}{d\tau }\Pi +F_{\bar{Q}}\frac{d\Pi }{d\tau }+\Lambda
\frac{d\Pi }{d\tau }\Bigg),
\end{eqnarray}

\begin{equation}
\frac{d\Lambda }{d\tau }=\left( -\frac{2}{9}M^{2}-F_{\bar{Q}}-\Lambda
\right) \Pi ,  \label{DFr3}
\end{equation}

\begin{equation}
\frac{d\Pi }{d\tau }=\frac{8}{3}h^{2}+\frac{4}{3}\frac{dh}{d\tau }%
-3h\Pi +\frac{1}{2}\Pi ^{2}.  \label{DFr4}
\end{equation}

From equations ~(\ref{DFr1}) and (\ref{DFr2}) we obtain the dynamical equation for the evolution of $h$ as 
\bea
\frac{dh}{d\tau}&=&-\frac{r+P}{4\Lambda}+\frac{1}{2\Lambda}\Bigg[-F_{\bar{T}}(r+P)-\frac{9}{4}F_{\bar{Q}}\Pi^2-\frac{1}{2}M^2\Pi^2\nonumber\\
&&-\frac{9}{4}\Lambda \Pi^2-\frac{7}{2}\Pi \frac{d\Lambda}{d\tau}+h\frac{d\Lambda}{d\tau}+\frac{2}{9}M^2\frac{d\Pi}{d\tau}+\Pi \frac{dF_{\bar{Q}}}{d\tau}\nonumber\\
&&+F_{\bar{Q}}\frac{d\Pi}{d\tau}+\Lambda \frac{d\Pi}{d\tau}\Bigg].
\eea

The effective energy density and pressure  become
\bea
r_{eff}&=&\frac{1}{\Lambda } \Bigg(F_{\bar{T}}(r+P)-\frac{F}{2}+\frac{9}{4}F_{%
\bar{Q}}\Pi {^{2}}+\frac{1}{4}M^{2}\Pi ^{2}+\frac{9}{8}{\Lambda }\Pi ^{2}\nonumber\\
&&+\frac{9}{4}\Pi\frac{d\Lambda }{d\tau } -3h\frac{d\Lambda }{d\tau }\Bigg),
\label{Dreff}
\eea
\bea
P_{eff}&=&\frac{1}{\Lambda }\Bigg(\frac{F}{2}+\frac{1}{4}M^{2}\Pi ^{2}+\frac{9%
}{8}\Lambda \Pi ^{2}+\frac{5}{4}\Pi\frac{d\Lambda }{d\tau } +2h\frac{d\Lambda
}{d\tau } \nonumber\\
&&-\frac{2}{9}M^{2}\frac{d\Pi }{d\tau }-\Pi\frac{dF_{\bar{Q}}}{d\tau }%
 -F_{\bar{Q}}\frac{d\Pi }{d\tau }-\Lambda \frac{d\Pi }{d\tau }\Bigg).
\eea

 The generalized energy balance equation (\ref{Fr11}) takes the dimensionless form
 \begin{multline}
     \frac{dr}{d\tau}+3h(r+P)-\frac{r}{\Lambda}\frac{d\Lambda}{d\tau}\\+2\Bigg[\frac{dr_{eff}}{d\tau}+3h\left(r_{eff}+P_{eff}\right)
 -\frac{r_{eff}}{\Lambda}\frac{d\Lambda}{d\tau}\Bigg]=0.
 \end{multline} 

\section{Warm inflation in Schr\"{o}dinger type $f(Q,T)$ gravity}\label{sect3}

In the present section, we explore the dynamics of the very early Universe within the framework of Schr\"{o}dinger-type $f(Q,T)$ gravity. We investigate the possibility that radiation-like matter is generated during the initial phase of cosmic evolution, which also describes accelerated expansion. This is achieved through the presence of the Schr\"{o}dinger vector field and the Lagrange multiplier. From a physical point of view, this scenario is similar to warm inflation. We begin by briefly reviewing standard warm inflation and then proceed to investigate warm inflation in Schrödinger-type $f(Q,T)$ gravity. Three models are considered: one assuming a constant non-metricity vector, which leads to an exact solution, one assuming a constant Lagrange multiplier, and a general model without imposing any restrictions on the model parameters, allowing the Lagrange multiplier to be dynamical.

\subsection{Brief review of standard warm inflation theory}

The inflationary description of the early Universe dynamics postulates a phase of accelerated expansion, driven by the inflaton field, which is usually considered a scalar field. In warm inflation, unlike in the cold inflation scenario, a radiation (matter) component coexists with the inflaton field, and interacts with it dynamically \cite{Fang, Be1,Be2, Be3}. As a result, the inflaton continuously transfers energy to the radiation field during inflation, thus generating the radiation/matter components without the need of a reheating period. The background evolution in warm inflation is governed by the modified Friedmann equations
\begin{equation}  \label{warmfriedmann}
3H^2 = {\frac{1 }{2}} \dot\phi^2 + V(\phi) + \rho_r\;,  2\dot{H} =
-\dot\phi^2 - {\frac{4 }{3}} \rho_r\,
\end{equation}
where $\phi$ is the inflaton field, $V(\phi)$ its potential, and $\rho_{r}$ the radiation energy density. Due to the inflaton-radiation interaction, energy is transferred from inflaton to radiation, leading to an increase in the energy density of the latter. The radiation creation process is described by the following conservation equations \cite{Fang, Be1,Be2, Be3}
\begin{eqnarray}  \label{warmconservation1}
\dot{\rho}_r + 3H (\rho_r + p_r) &=& \Gamma \dot\phi^2\;, \\
\dot{\rho}_\phi + 3H(\rho_\phi + p_\phi) &=& -\Gamma \dot\phi^2\;,
\label{warmconservation}
\end{eqnarray}
where $\Gamma$ is the dissipation coefficient. Equation \eqref{warmconservation} also
gives the equation of motion for $\phi$, which is obtained as
\begin{equation}  \label{warmeom}
\ddot{\phi} + 3H (1+\mathcal{Q}) \; \dot{\phi} + V^{\prime }(\phi) =0\;,
\end{equation}
where we have introduced the parameter, $\mathcal{Q}=\Gamma/3H$, representing the ratio of the dissipation coefficient
and the expansion rate. The notation $\mathcal{Q}$ shall not be confused with the nonmetricity $Q$.

In warm inflation, the slow-roll approximation remains applicable. To maintain a quasi-de Sitter expansion, the Hubble parameter must vary slowly over one Hubble time. This condition is imposed in equations ~(\ref{warmfriedmann}) via the smallness of the first slow-roll parameter $\epsilon_1$, defined according to
$\epsilon_1 = - \dot{H}/ H^2$ \cite{101,102}.

During this phase, inflaton energy density dominates over radiation, i.e. $\rho_\phi \gg \rho_r$, and its kinetic energy is negligible compared to its potential, so  $\rho_\phi \simeq V(\phi)$. Moreover, it is assumed that the radiation production is quasi-stable, so that $\dot{\rho_r} \ll H\rho_r$ and $\dot{\rho_r} \ll \Gamma \dot\phi^2$. Under these assumptions, and using \eqref{warmfriedmann}, \eqref{warmconservation}, and \eqref{warmeom}, the warm inflationary dynamics reduce to the following system
\begin{eqnarray}
&&3H^2 \simeq V(\phi)\;,  \label{vfriedmann} \\
&&\dot\phi \simeq - {\frac{V^{\prime }(\phi) }{3H(1+\mathcal{Q})}}\;,  \label{dotphi}
\\
&& \rho_r = k_B T^4 = {\frac{\Gamma }{4H}} \dot\phi^2\;,
\label{radiationtemp}
\end{eqnarray}
where $T$ is the temperature of the radiation fluid, $k_B=\pi^2 g_\star / 30$ is the Stefan-Boltzmann constant, and $g_\star$ denotes the number of degrees of freedom of the radiation fluid.

An important cosmological parameter, the number of e-folds is defined according to the relation
\begin{equation}\label{efold}
N = \int_{t_\star}^{t_e} H dt = \int_{\phi_\star}^{\phi_e} {\frac{H }{%
\dot\phi}} \; d\phi = - \int_{\phi_\star}^{\phi_e} (1+\mathcal{Q}) {\frac{V(\phi) }{%
V^{\prime }(\phi)}} \; d\phi\;,
\end{equation}
where the subscripts $``e"$ and $``\star$" denote the values of $t$ at the end of inflation and at the horizon crossing, respectively. The last equality in Eq. ~(\ref{efold}) is obtained by using Eqs.~\eqref{vfriedmann} and \eqref{dotphi}, respectively. The smallness of the slow-roll parameters guarantees that the Universe had in its early phases a quasi-exponential accelerated expansionary evolution, and that it remained in this phase for a number of e-folds that can solve the problems of the hot Big Bang model.

\subsection{Warm inflation in Schr\"{o}dinger type $f(Q,T)$ theory}

As a first cosmological application of the Schr\"{o}dinger-type $f(Q,T)$ theory, we investigate the possibility that radiation in the very early Universe is generated purely from the underlying spacetime geometry. That is, we  consider an alternative inflationary model, in which both the expansion dynamics and matter creation arise without invoking an inflaton field. Instead, we adopt a warm inflationary scenario where radiation is produced during an accelerated expansion phase, sourced by the Schr\"{o}dinger vector field and the associated Lagrange multiplier. Since radiation satisfies $p=\frac{\rho}{3}$, the trace of the energy-momentum tensor vanishes, i.e. $T=0$. Therefore, we can neglect any $T$-dependence in the action and the resulting field equations. Moreover, we do not assume that the mass $m$ of the Schr\"{o}dinger vector field is negligible, and thus we generally take $m \neq 0$. For simplicity, we adopt the simplest form of the function $f$, given by
\be
f(Q)=\alpha Q, 
\ee
where $\alpha $ is a constant. In the dimensionless representation, we obtain $F\left(\bar{Q}\right)=\alpha \bar{Q}$. Moreover, since we have $\bar{Q}=(9/4)\Pi^2$, we can write, $F\left(\bar{Q}\right)= \alpha(9/4)\Pi^2$.

As a first step in our analysis, we reformulate the balance equation (\ref{Fr11}) as
\begin{equation}
\dot{\rho}+4H\rho =\Gamma ,  \label{Fr12}
\end{equation}
where
\begin{equation}
\Gamma =\rho \frac{\dot{\lambda}}{\lambda }-2\left[ \dot{\rho}
_{eff}+3H\left( \rho _{eff}+p_{eff}\right) -\rho _{eff}\frac{\dot{\lambda}}{ \lambda }\right] ,  \label{Fr13}
\end{equation}
If $\Gamma (t)>0$, matter creation can take place during the cosmological evolution. If $\Gamma (t)\equiv 0$, $\forall\  t\geq 0$, then the total
energy-matter content of the Universe is conserved, and no radiation or matter creation is possible.

\subsubsection{Warm inflation in the presence of a constant non-metricity vector - an exact solution of the field equations}

As a first example of an inflationary type model in the Schr\"{o}dinger type $f(Q,T)$ gravity, we will consider the case in which the non-metricity vector $\Pi$ is constant, $\Pi=\Pi_0={\rm constant}$. With this choice the dimensionless evolution equation ~(\ref{DFr4}) reduces to
\be
\frac{dh}{d\tau}-\frac{9}{4}\Pi_0h+2h^2+\frac{3}{8}\Pi_0^2=0,
\ee 
which admits the general solution
\be
h(\tau)=\frac{9}{16}\Pi_0+\frac{\sqrt{33}}{16}\Pi_0\tanh \left[\frac{\sqrt{33}}{8}\Pi_0\left(\tau-\tau_0\right)\right],
\ee
where $\tau_0$ is an arbitrary integration constant, which can be taken to be zero, i.e., $\tau_0=0$. Eq.~(\ref{DFr3}) takes the form
\be
\frac{d\Lambda}{d\tau}=\left(-\frac{2}{9}M^2-\alpha -\Lambda\right)\Pi_0,
\ee
and it has the solution 
\be
\Lambda (\tau)=\frac{1}{9}\left[\left(-2M^2-9\alpha\right)+C_1e^{-\Pi_0 \tau}\right],
\ee
where $C_1$ is an arbitrary integration constant. The evolution of the matter density can be obtained from equation ~(\ref{DFr1}), and it is given by
\bea
r&=&\frac{9}{128} \Pi_0^2 e^{-\Pi_0 t} \left[11 C_1-3 \left(9 \alpha +2
   M^2\right) e^{\Pi_0 t}\right]\nonumber\\
   &&+\frac{11}{128} \Pi_0^2 e^{-\Pi_0 t} \tanh
   ^2\left(\frac{1}{8} \sqrt{33} \Pi_0 t\right) \nonumber\\
  &&\times  \left[9 C_1-\left(9 \alpha +2 M^2\right)
   e^{\Pi_0 t}\right]\nonumber\\
   &&+\frac{3}{64} \sqrt{33} \Pi_0^2 e^{-\Pi_0 t} \tanh
   \left(\frac{1}{8} \sqrt{33} \Pi_0 t\right)\nonumber\\
   &&\times  \left[C_1-\left(9 \alpha +2 M^2\right)
   e^{\Pi_0 t}\right].
\eea

\begin{figure}[htp]
\includegraphics[width=8.0cm]{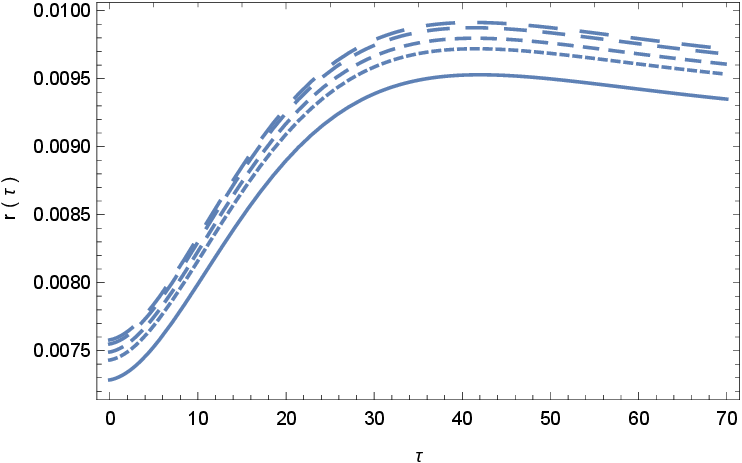}
\caption{Variation of the matter energy density during the warm inflationary phase of evolution of the Universe in the Schr\"{o}dinger type $f(Q,T)$ theory, for different values of the constant non-metricity vector $\Pi_0$: $\Pi_0 =0.05$ (solid curve), $\Pi_0 =0.0.0505$ (short dashed curve), $\Pi_0 =0.0507$ (dashed curve), $\Pi_0 =0.0509$ (long dashed curve), and $\Pi_0 =0.051$ (ultra-long dashed curve). The numerical values of the free parameters of the constant non-metricity model have been fixed to $C_1=2$, $M=0.9$, and $\alpha =-0.9$, respectively.}
\label{fig0}
\end{figure}

The behavior of the matter energy density as a function of $\tau$ is illustrated in Fig. ~\ref{fig0}, for various values of $\Pi_0$ and fixed values of the parameters $C_1,M$ and $\alpha$.

For the adopted values of the model parameters the matter energy density has a finite value at the beginning of the cosmological expansion. The energy density increases for an initial period $\tau \in \left[0,\tau_{max}\right]$, and reaches a maximum value, followed by a decreasing trend, due to the expansion of the Universe. Hence, this model describes a period of matter/energy creation in the early Universe, the creation process being triggered, and supported, by the non-metricity of the spacetime. The variation of the energy density is strongly dependent on the numerical values of the constant non-metricity vector $\Pi_0$. As time tends to large values, the Hubble function settles at a constant value, $\lim_{\tau \rightarrow \infty}h(\tau)=\left(9+\sqrt{33}\right)\Pi_0/16$. In the same limit, the Lagrange multiplier also tends to a constant value, $\lim_{\tau \rightarrow \infty}\Lambda(\tau)=\left(-2M^2-9\alpha\right)/9$, a value which is independent on $\Pi_0$.

\subsubsection{Warm inflationary models with a constant Lagrange multiplier}

Next, we consider the scenario in which the Lagrange multiplier $\lambda$ is approximately constant, i.e. $\lambda \simeq$ constant. Under this assumption, and using equation \eqref{Fr3} along with the choice $f_{Q}=\alpha$, we obtain
\begin{equation}
\lambda =\lambda _{0}=-\kappa ^{2}\alpha -\frac{2}{9}m^{2}.
\end{equation}

For a negligibly small $m^{2}$, it follows that the Lagrange multiplier is proportional to the constant, $\alpha $, present in the action. The non-metricity scalar is $Q=-\frac{9}{4} \pi_{\alpha} \pi^{\alpha}$, which in cosmology takes the form $Q=-\frac{9}{4}\left( \pi (-\pi) \right)=+\frac{9}{4} \pi^2$.

The expressions for the effective energy and pressure take the form
\begin{eqnarray}
\rho _{eff} &=&-\frac{\kappa ^{2}\alpha }{2}Q+\left( \frac{9}{4}\kappa
^{2}\alpha +\frac{1}{4}m^{2}+\frac{9}{8}{\lambda }_{0}\right) \pi ^{2} \\
&=&\left( \frac{9}{8}\kappa ^{2}\alpha +\frac{1}{4}m^{2}-\frac{9}{8}\kappa
^{2}\alpha -\frac{1}{4}m^{2}\right) \pi ^{2}\equiv 0 \; \;,
\end{eqnarray}
and
\begin{equation}
\begin{aligned}
p_{eff}=&\frac{\kappa^2\alpha}{2} \left( \frac{9}{4} \pi^2 \right) +\frac{1}{4} m^2 \pi^2+\frac{9}{8} \lambda_0 \pi^2 - \frac{2}{9} m^2 \dot \pi \\
&- \kappa^2 \alpha \dot \pi -\lambda_0 \dot \pi\equiv 0, 
\end{aligned}
\end{equation}
respectively. Hence in this case the system of field equations of the Schr\"{o}dinger type $f(Q,T)$ gravity reduce to the case of the standard general relativity, since the contributions of the vector field identically vanish.  Hence, in order to construct a warm inflationary model in this scenario, the presence of another dynamical quantity, like, for example, a scalar field, is required in the theoretical framework. 

\subsubsection{Warm inflation in the presence of a dynamical $\lambda$}

We will consider now the full set of cosmological field equations, by assuming that the Lagrange multiplier is a function  of time. The evolution equations describing the dynamics of the very early Universe in the presence of radiation production are then given by
\bea\label{f1}
&& \left[1-\frac{2}{27} \left(9 \alpha +2 M^2+9 \Lambda \right)\right]\frac{dh}{d\tau}\nonumber\\
&&+\frac{1}{216}
   \Bigg[48 h \Pi  \left(9 \alpha +2 M^2+9 \Lambda \right)-32 h^2 \left(9 \alpha
   +2 M^2+9 \Lambda \right)\nonumber\\
 &&  -3 \Pi ^2 \left(144 \alpha +14 M^2+63 \Lambda
   \right)+\frac{72 r}{\Lambda }\Bigg]=0,
\eea 
\begin{equation}
\frac{d\Lambda }{d\tau }=\left( -\frac{2}{9}M^{2}-\alpha-\Lambda
\right) \Pi ,  \label{f2}
\end{equation}
\begin{equation}
\frac{d\Pi }{d\tau }=\frac{8}{3}h^{2}+\frac{4}{3}\frac{dh}{d\tau }%
-3h\Pi +\frac{1}{2}\Pi ^{2},  \label{f3}
\end{equation}
\bea\label{f4}
 \frac{dr}{d\tau}+4hr-\frac{r}{\Lambda}\frac{d\Lambda}{d\tau}&+&2\Bigg[\frac{dr_{eff}}{d\tau}+3h\left(r_{eff}+P_{eff}\right)\nonumber\\
 &-&\frac{r_{eff}}{\Lambda}\frac{d\Lambda}{d\tau}\Bigg]=0.
 \eea 

\begin{figure}[htp]
\includegraphics[width=8.0cm]{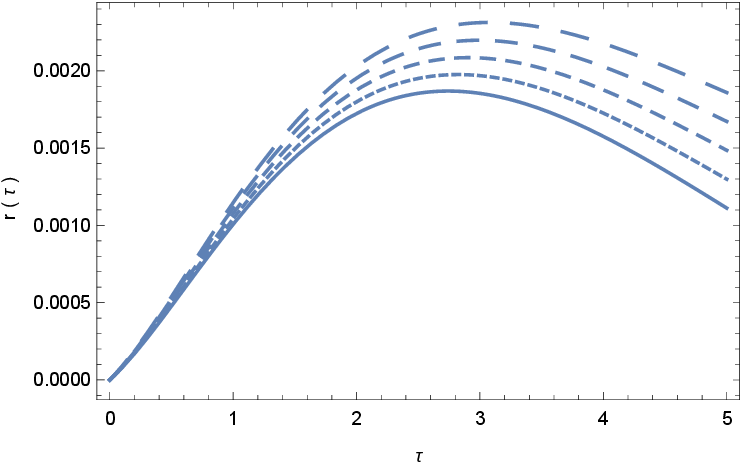}
\caption{Variation of the radiation energy density during the warm inflationary phase of evolution of the Universe in the Schr\"{o}dinger type $f(Q,T)$ theory, for various values of $\alpha$: $\alpha =6$ (solid curve), $\alpha =7$ (short dashed curve), $\alpha =8$ (dashed curve), $\alpha =9$ (long dashed curve), and $\alpha =10$ (ultra-long dashed curve). The numerical values of $M$ has been fixed to  $M^2=1$. The initial value of the time variable was taken as $\tau _0=10^{-10}$. The initial conditions used to numerically integrate the system of evolution equations are $h\left(\tau_0\right)=0.11$, $r\left(\tau_0\right)=10^{-10}$,  $\Pi \left(\tau_0\right)=0.01$, and $\Lambda \left(\tau_0\right)=150$, respectively.  }
\label{fig6}
\end{figure}

\begin{figure}[htp]
\includegraphics[width=8.0cm]{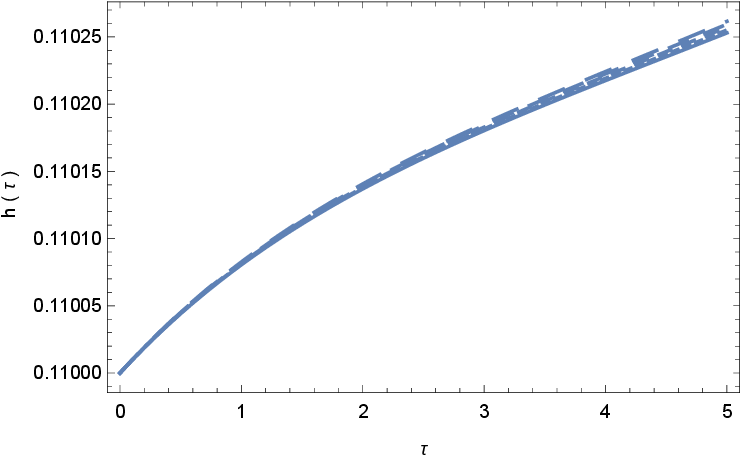}
\caption{Variation of the Hubble function during the warm inflationary phase of evolution of the Universe in the Schr\"{o}dinger type $f(Q,T)$ theory, for various values of $\alpha$: $\alpha =6$ (solid curve), $\alpha =7$ (short dashed curve), $\alpha =8$ (dashed curve), $\alpha =9$ (long dashed curve), and $\alpha =10$ (ultra-long dashed curve). The numerical values of $M$ has been fixed to  $M^2=1$. The initial value of the dimensionless time variable $\tau$  was taken as $\tau _0=10^{-10}$. The initial conditions used to numerically integrate the system of evolution equations are $h\left(\tau_0\right)=0.11$, $r\left(\tau_0\right)=10^{-10}$,  $\Pi \left(\tau_0\right)=0.01$, and $\Lambda \left(\tau_0\right)=150$, respectively.  }
\label{fig7}
\end{figure}

The system of equations (\ref{f1})-(\ref{f4}) must be integrated with the initial conditions $h(0)=h_0$, $r(0)=0$, $\Lambda (0)=\Lambda_0$, and $\Pi (0)=\Pi_0$, respectively. The variation of the energy density of the radiation for the general Schr\"{o}dinger $f(Q,T)$ type warm inflationary model is presented in Fig.~\ref{fig6}. 
The energy density of the radiation fluid increases during the expansion of the Universe, and it reaches a maximum value, $r_{max}$, at a finite time, $\tau=\tau_{max}$. The maximum value of the radiation fluid, as well as its temporal variation depends strongly on the model parameter, $\alpha$, which describes the $f(Q,T)$ dependence of the model. The variation of the Hubble function is shown in Fig.~\ref{fig7}.  

\begin{figure*}[htp]
\includegraphics[width=8.0cm]{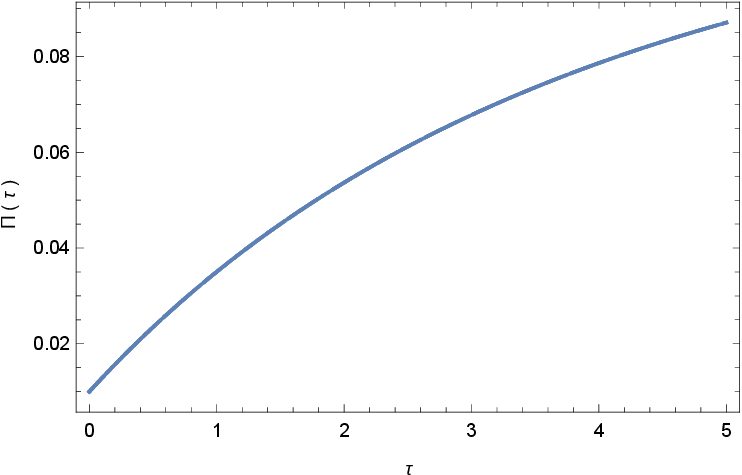}
\includegraphics[width=8.0cm]{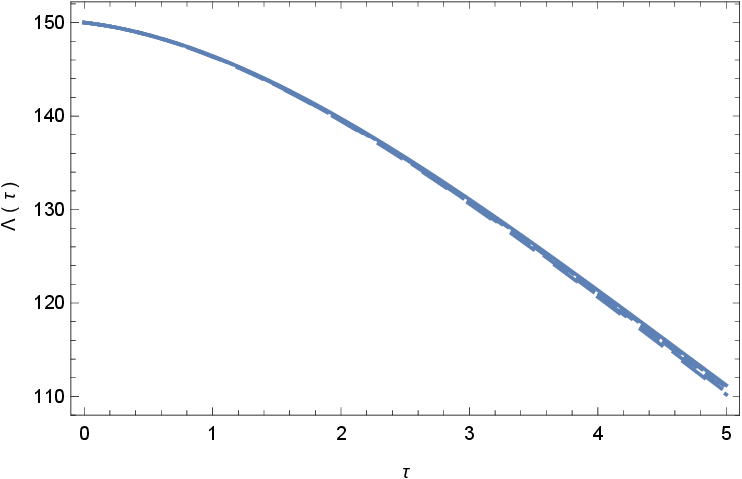}
\caption{Time variations Schrödinger\"{o}dinger vector (left panel) and of the Lagrange multiplier $\lambda$ (right panel)  during the warm inflationary phase of evolution of the Universe in the Schr\"{o}dinger type $f(Q,T)$ gravity, for various values of $\alpha$: $\alpha =6$ (solid curve), $\alpha =7$ (short dashed curve), $\alpha =8$ (dashed curve), $\alpha =9$ (long dashed curve), and $\alpha =10$ (ultra-long dashed curve). The numerical values of $M$ has been fixed to  $M^2=1$. The initial value of the dimensionless time variable $\tau$ was taken as $\tau _0=10^{-10}$. The initial conditions used to numerically integrate the system of evolution equations are $h\left(\tau_0\right)=0.11$, $r\left(\tau_0\right)=10^{-10}$,  $\Pi \left(\tau_0\right)=0.01$, and $\Lambda \left(\tau_0\right)=150$, respectively.  }
\label{fig8}
\end{figure*}

The Hubble function monotonically increases during the warm inflationary phase, and its evolution shows a weak dependence on the numerical values of $\alpha$. The evolution of Schr\"{o}dinger vector and the Lagrange multiplier with time is depicted in Fig.~\ref{fig8}. The time evolutions of these quantities are effectively independent of the numerical values of $\alpha$. While the Schr\"{o}dinger vector is increasing during the warm inflationary period, the Lagrange multiplier shows a significant decrease, indicating that it is the main source of radiation and also of geometry production, leading to an increase of the non-Riemannian effects in the early Universe.  

\section{Late Universe observational tests of the Schr\"{o}dinger type $f(Q,T)$ cosmological models}\label{sect4}

We consider now another possibility for testing  the Schr\"{o}dinger type $f(Q,T)$ gravity theory through the analysis of the concordance of the theoretical predictions of a specific model, and the observations.  To perform the comparison with the observational data we consider the late evolution of the Universe, which we assume to be described by the system of generalized Friedmann equations (\ref{DFr1})-(\ref{DFr4}). We assume that the matter content of the Universe consists of pressureless dust, and thus we take $P=0$ in the cosmological evolution equations. In order to obtain specific cosmological models we need to fix the functional form of $f(Q,T)$. In the following We will consider the simplest possible case, in which the action of the gravitational-geometric sector of the theory is given by $f(Q,T)=\left(\alpha /\kappa ^2H_0^2\right)T+\left(4\beta/9H_0^2\right)Q$. The nonmetricity scalar $Q$ is given by $Q=(9/4)\Pi^2$, while for the trace of the matter energy-momentum tensor we take $T=-\rho$. We also introduce as the independent variable the redshift $z$, defined as $1+z=1/a$, which allows to replace the derivative with respect to time with the derivative with respect to the redshift according to the rule
\be
\frac{d}{d\tau}=-(1+z)h(z)\frac{d}{dz}.
\ee 

\subsection{Specific cosmological model: $f(Q,T)=\left(\alpha /\kappa ^2H_0^2\right)T+\left(4\beta/9H_0^2\right)Q$}

As an example of a specific cosmological model in the Schr\"{o}dinger type $f(Q,T)$ gravity we adopt for the Lagrangian density $f(Q,T)$ the simple form
\bea
f(Q,T)&=&\left(\frac{\alpha} {\kappa ^2H_0^2}\right)T+\left(\frac{4\beta}{9H_0^2}\right)Q, \\
F\left(\bar{Q}, \bar{T}\right)&=&\alpha \bar{T}+\frac{4}{9}\beta \bar{Q},
\eea  
where $\alpha$ and $\beta $ are constants. For this choice of $f$ we obtain  $F_{\bar{T}}=\alpha$, and $F_{\bar{Q}}=4\beta/9$, respectively. 

The system of the generalized Friedmann equations (\ref{DFr1})-(\ref{DFr4}) takes the form
\begin{eqnarray}\label{87}
\hspace{-0.5cm}\frac{dh}{dz} &=&-\frac{9 \Pi ^2 \left(4 \beta +2 M^2+9 \Lambda \right)}{16 (3
   \alpha +1) (z+1) h \left(8 \beta +4 M^2-9 \Lambda
   \right)}\nonumber\\
 &&  +\frac{h \left[16(1+3\alpha)\left(2\beta +M^2\right)+9(6\alpha-1)\Lambda \right]}{2 (3
   \alpha +1) (z+1) \left(8 \beta +4 M^2-9 \Lambda
   \right)}\nonumber\\
   &&-\frac{3 (6 \alpha +1) \Pi  \left(4 \beta +2 M^2+9
   \Lambda \right)}{2 (3 \alpha +1) (1+z) \left(8 \beta +4 M^2-9
   \Lambda \right)},
\end{eqnarray}%
\begin{eqnarray}\label{88}
\hspace{-0.5cm}\frac{d\Pi }{dz} &=&\frac{\Pi ^2 \left[2(12\alpha+1)\left(2\beta +M^2\right)-9(6\alpha +5)\Lambda \right]}{4 (1+3 \alpha)
   (z+1) h \left(8 \beta +4 M^2-9 \Lambda \right)}\nonumber\\
   &&+\frac{18 (6
   \alpha +1) h \Lambda }{(3 \alpha +1) (z+1) \left(8 \beta +4
   M^2-9 \Lambda \right)}\nonumber\\
   &&+\frac{\Pi  \left[4(3\alpha+2)\left(2\beta +M^2\right)-21(9\alpha +5)\Lambda
   \right]}{(3 \alpha +1) (z+1) \left(8 \beta +4 M^2-9 \Lambda
   \right)}, 
\end{eqnarray}
\be\label{89}
   \frac{d\Lambda}{dz}= \frac{\Pi  \left(4 \beta +2 M^2+9 \Lambda
   \right)}{9 (1+z) h}.
   \ee
 
 The system of differential equations (\ref{87})-(\ref{89}) must be integrated with the initial conditions $h(0)=1$, $\Pi (0)=\Pi_0$ and $\Lambda (0)=\Lambda_0$, respectively. The variation of the matter energy density is given by 
 \bea
 r&=&\frac{6 h^2 \Lambda }{1+3 \alpha }-\frac{2 h \Pi  \left(4 \beta +2 M^2+9 \Lambda \right)}{3 (1+3
   \alpha )}\nonumber\\
  && +\frac{ \left(4 \beta +2 M^2+9 \Lambda \right)\Pi
   ^2}{4 (1+3 \alpha)}.
  \eea

\subsection{Methodology and Datasets}

To constrain the parameters of the Schr\"{o}dinger-type $f(Q,T)$ gravity model, we first solve the system of differential equations that represent the Schr\"{o}dinger-type $f(Q,T)$ model, given by the system of equations (\ref{87})--(\ref{89}). These coupled nonlinear ordinary differential equations describe the redshift evolution of the dimensionless Hubble parameter $h(z)$.

To solve the system numerically, we use the \textit{solve\_ivp} function from the \textit{SciPy} library~\citep{Virtanen2020}, using the \textit{Radau} implicit integration method, which is particularly efficient and stable for stiff systems of differential equations. The integration is performed over the redshift interval $z \in [0,3]$, with the initial conditions $h(0) = 1$, $\Pi(0) = \Pi_0$, and $\Lambda(0) = \Lambda_0,$ and the parameter vector defined as $\Theta = (H_0,\, \Lambda_0,\, \Pi_0,\, \alpha,\, \beta,\, M,\, r_d)$. 

During the analysis, we impose the following uniform (flat) priors on the model parameters: $H_0 \in [50.0,100.0], \quad
\Lambda_0 \in [-2.0,0.0], \quad
\Pi_0 \in [0.0,1.0], \quad
\alpha \in [-3.0,0.0], \quad
\beta \in [2.0,0.0], \quad
M \in [0.8,1.2], \quad
r_d \in [100.0,300.0], \quad
\Omega_m0 \in [0.0,1.0].$ The solver uses adaptive step sizes with relative and absolute tolerances set to $10^{-3}$ and $10^{-6}$ to maintain numerical stability and accuracy. From the solutions for $\Lambda(z)$, $\Pi(z)$, and $h(z)$, the Hubble expansion rate is obtained as $H(z) = H_0\, h(z).$

Once the numerical solutions is obtained, we compare the theoretical predictions with observational data to constrain the model parameters. For this, we use the \textsc{Dynesty} package~\cite{Speagle2020}, which implements the Nested Sampling algorithm proposed by \cite{Skilling2004,Skilling2006}. Nested Sampling is a Bayesian technique designed to efficiently estimate both the posterior distributions of parameters and the Bayesian evidence~($\mathcal{Z}$), which quantifies the overall support for a given model. Unlike the traditional Markov Chain Monte Carlo methods that sample points proportional to the posterior, Nested Sampling explores the parameter space by successively shrinking prior volumes around regions of higher likelihood, allowing for simultaneous estimation of the evidence and the posterior.

The Bayesian evidence is defined as the integral of the likelihood function $\mathcal{L}(\Theta)$ over the parameter space, weighted by the prior probability $\pi(\Theta)$:
\[
\mathcal{Z} = \int \mathcal{L}(\Theta)\, \pi(\Theta)\, d\Theta,
\]
where $\Theta$ denotes the set of model parameters. Nested Sampling transforms this multidimensional integral into a 1D  integral over the prior volume $X(\lambda)$ enclosed by the iso-likelihood contour $\mathcal{L}(\Theta) > \lambda$:
\[
\mathcal{Z} = \int_0^1 \mathcal{L}(X)\, dX.
\]
This formulation allows for a more efficient numerical evaluation of $\mathcal{Z}$, particularly for complex or multi-modal likelihood surfaces.

The difference in logarithmic evidence between two competing models, $\Delta \ln \mathcal{Z}$, is used to quantify the relative statistical preference. A positive $\Delta \ln \mathcal{Z}$ indicates support for the model with the higher evidence. To interpret the strength of this evidence, we follow the revised Jeffreys’ scale~\citep{Kass1995}, which classifies the evidence as follows:
\[
\begin{cases}
0 \leq |\Delta \ln \mathcal{Z}| < 1, & \text{inconclusive or weak support,} \\
1 \leq |\Delta \ln \mathcal{Z}| < 3, & \text{moderate evidence,} \\
3 \leq |\Delta \ln \mathcal{Z}| < 5, & \text{strong evidence,} \\
|\Delta \ln \mathcal{Z}| \geq 5, & \text{decisive evidence.}
\end{cases}
\]

Furthermore, we also use several statistical metrics to compare the performance of the $\Lambda$CDM and Schr\"{o}dinger-type $f(Q,T)$ model. These include the minimum Chi-squared value ($\chi^2_{\mathrm{min}}$), the reduced Chi-squared statistic ($\chi^2_{\mathrm{red}}$), the Akaike Information Criterion (AIC), and the Bayesian Information Criterion (BIC) \cite{Liddle,AIC1,AIC2,AIC3,BIC1}. We also compute their respective differences, $\Delta$AIC and $\Delta$BIC, to quantify the relative preference between the models. The interpretation of $\Delta$AIC and $\Delta$BIC$ $ follows the Jeffreys' scale~\cite{Jeffreys}, where smaller values indicate a more favored model. Specifically:
\[
\begin{cases}
|\Delta \text{AIC}| \leq 2 & \text{: Models are statistically comparable},\\
4 \leq |\Delta \text{AIC}| < 10 & \text{: Considerably less support for the model},\\
|\Delta \text{AIC}| \geq 10 & \text{: Strongly disfavored},\\[6pt]
|\Delta \text{BIC}| \leq 2 & \text{: Weak evidence against the model},\\
2 < |\Delta \text{BIC}| \leq 6 & \text{: Moderate evidence against the model},\\
|\Delta \text{BIC}| > 6 & \text{: Strong evidence against the model}.
\end{cases}
\]
In addition, we evaluate the corresponding $p$-values for each model to assess the statistical significance of the fits \cite{Ioannidis2018}. This approach provides a robust framework for both parameter estimation and model comparison using various statistical metrics within the Schr\"{o}dinger-type $f(Q,T)$ gravity model.

For the posterior visualization, we use the \textsc{GetDist} package~\citep{Lewis2025} to generate the marginalized parameter contours. In our analysis, we use the Cosmic Chronometers, Type Ia Supernovae, and Baryon Acoustic Oscillation datasets. Below, we provide detailed descriptions of each observational dataset used in this work.
\begin{itemize}
\item \textbf{Cosmic Chronometers :} We use Hubble parameter measurements obtained from the differential age method. This approach studies massive galaxies that have evolved passively since their formation at redshifts around ( z $\sim$ 2–3 ). By comparing the change in redshift with the change in age of these galaxies, the Hubble parameter can be estimated directly and without assuming a specific cosmological model \cite{Jimenez}. In this analysis, we use 15 measurements spanning the redshift range $0.17 \leq z \leq 1.96$ reported in \cite{Moresco2016,Moresco2018,Moresco2020}, as these include the full covariance matrix accounting for both statistical and systematic uncertainties \cite{Moresco2012,Moresco2015}. The corresponding likelihood used in this work is taken from the implementation provided by Moresco on his GitLab repository\footnote{\protect\url{https://gitlab.com/mmoresco/CCcovariance}}
\item \textbf{Unanchored SNe Ia:} We also use the Pantheon$^{+}$ (PP) compilation, which consists of 1,701 light curves from 1,550 Type Ia supernovae (SNe Ia) covering the redshift range $0.001 \leq z \leq 2.26$ \cite{Brout2022}. In our analysis, we consider 1,590 light curves, excluding those at $z < 0.01$ due to significant systematic uncertainties associated with peculiar velocities. We use the PP CosmoSIS likelihood in our analysis, provided in the following GitHub repository\footnote{\protect\url{[https://github.com/PantheonPlusSH0ES/DataRelease}}. We marginalize over the nuisance parameter $\mathcal{M}$ ; for additional details, see Eqs.~(A9–A12) of \cite{Goliath2001}.

\item \textbf{Baryon Acoustic Oscillation :} Finally, we use the recent Baryon Acoustic Oscillation (BAO) measurements from the Dark Energy Spectroscopic Instrument (DESI) Data Release~2 (DR2)\footnote{\protect\url{https://github.com/CobayaSampler/bao_data}}~\cite{Karim2025}. These measurements are reported at the effective redshift \( z_{\mathrm{eff}} \), which corresponds to the redshift with the maximum statistical weight of each sample. To constrain the cosmological parameters using the DESI~DR2 measurements, we compute the following distance measures: the Hubble distance $D_H(z) = \frac{c}{H(z)}$, the comoving angular diameter distance $D_M(z) = c \int_0^z \frac{dz'}{H(z')}$, and the volume-averaged distance $D_V(z) = \left[z\,D_M^2(z)\,D_H(z)\right]^{1/3}.$ We then compare these quantities with the observational measurements expressed as \( D_H(z)/r_d \), \( D_M(z)/r_d \), and \( D_V(z)/r_d \), where \( r_d \) denotes the sound horizon. The sound horizon is defined as $r_d = \int_{z_d}^{\infty} \frac{c_s(z)}{H(z)}\,dz,$
where \( c_s(z) \) is the sound speed of the photon baryon fluid. In the standard flat \( \Lambda \)CDM model, the sound horizon is \( r_d = 147.09 \pm 0.26~\mathrm{Mpc} \)~\cite{Aghanim2020}. In our analysis, we treat $r_d$ as a free parameter \cite{Pogosian2020,Jedamzik2021,Pogosian2024,Lin2021,Vagnozzi2023}.\\
\end{itemize}
To constrain the parameters of both cosmological models, we minimize the total chi-squared function, which can be written as $\chi^2_{\mathrm{tot}} = \chi^2_{\mathrm{BAO}} + \chi^2_{\mathrm{SNe,Ia}} + \chi^2_{\mathrm{CC}},$ which is equivalent to maximizing the total likelihood $\mathcal{L}_{\mathrm{tot}} \propto e^{-\chi^2{\mathrm{tot}}/2}.$

\begin{figure*}
\centering
\includegraphics[scale=0.90]{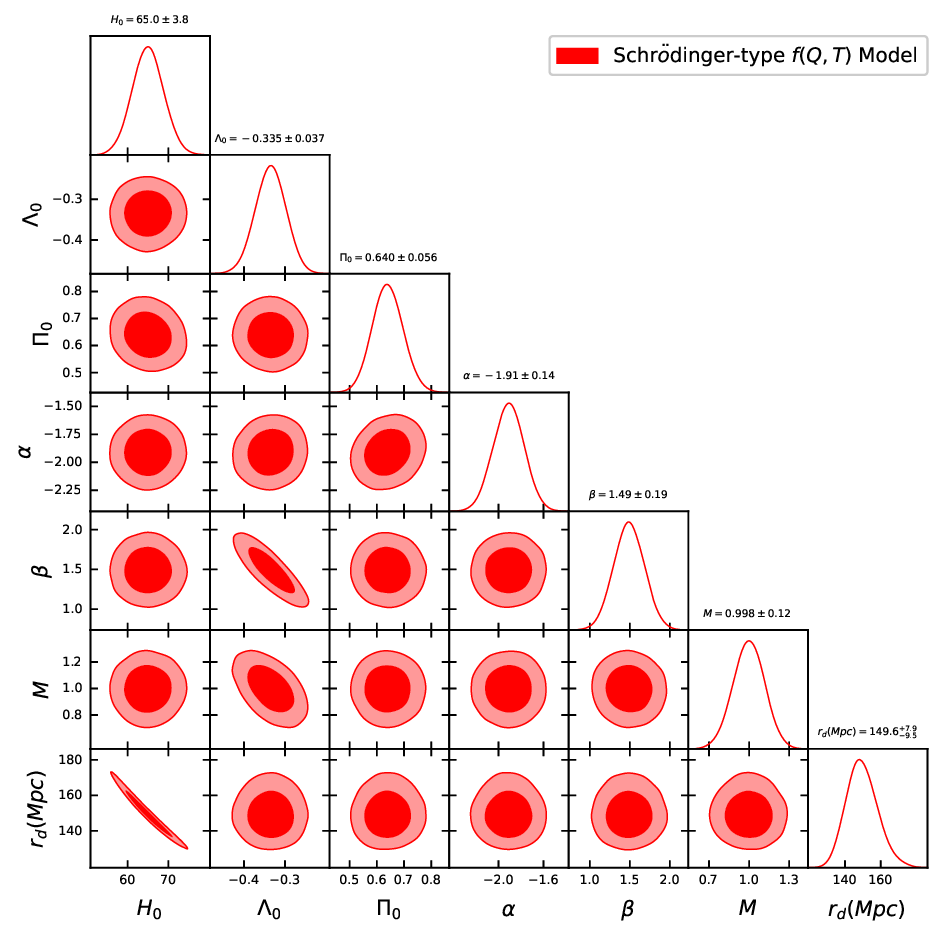}
\caption{The posterior distributions of the parameters of the Schr\"{o}dinger-type $f(Q,T)$ Model with 68\% ($1\sigma$) and 95\% ($2\sigma$) confidence levels.}\label{fig_12}
\end{figure*}
\begin{table*}
\centering
\begin{tabular}{lcc}
\hline
\textbf{Parameter} & \textbf{$\Lambda$CDM} & \textbf{Schr\"{o}dinger-type $f(Q,T)$ Model} \\
\hline
$H_0$ & $67.8 \pm 3.7$  & $65.0 \pm 3.8$  \\
$\Omega_{mo}$ & $0.3092 \pm 0.0086$  & --- \\
$\Lambda_0$ &  --- & $-0.335 \pm 0.037$ \\ 
$\Pi_0$ & --- & $0.640 \pm 0.056$  \\
$\alpha$ & --- & $-1.91 \pm 0.14$  \\
$\beta$ &  --- & $1.49 \pm 0.19$  \\
$M$ & ---  & $0.998 \pm 0.12$  \\
$r_d$ & $148.5 \pm 7.5$ & $149.5_{-9.8}^{+7.5}$  \\
$\chi^2_{min}$ & 1574.88 & 1535.62  \\
$\chi^2_{red}$ & 0.975 & 0.953  \\
AIC & 1580.88 & 1549.62  \\
$\Delta$AIC & 0 & -31.26  \\
BIC & 1597.04 & 1587.34  \\
$\Delta$BIC & 0 & -9.70 \\
P-value & 0.758 & 0.909  \\
$|\Delta \ln \mathcal{Z}_{\Lambda\mathrm{CDM}, \mathrm{Model}}|$ & 0 & 8.13 \\
\hline
\end{tabular}
\caption{The numerical results for the $\Lambda$CDM and Schr\"{o}dinger-type $f(Q,T)$ Model obtained using the MCMC analysis, and statistical metrics.}
\label{tab_1}
\end{table*}


\subsection{Comparing the Schr\"{o}dinger-type $f(Q,T)$ and the $\Lambda$CDM models}

In this subsection, we compare the Schr\"{o}dinger-type $f(Q,T)$ model with the standard $\Lambda$CDM cosmology. We plot the Hubble function and its residuals with respect to redshift after obtaining the corresponding mean values of the model parameters from the MCMC analysis. This comparison allows us to assess the compatibility of the $f(Q,T)$ model with the $\Lambda$CDM model and the CC measurements \cite{Moresco2012,Moresco2015,Moresco2016}.\\

\subsubsection{Evolution of the Hubble Parameter and Hubble Residual:}
To compare the Schr\"{o}dinger-type $f(Q,T)$ model with the $\Lambda$CDM model and the observational Hubble data, we first compute the Hubble function for the $\Lambda$CDM model using
\begin{equation}
H_{\Lambda \text{CDM}}(z) = H_0 \sqrt{\Omega_{m0} (1 + z)^3 + (1 - \Omega_{m0})},
\end{equation}
where $H_0 = 67.8~\mathrm{km\,s^{-1}\,Mpc^{-1}}$ and $\Omega_{m0} = 0.309$.

For the Schr\"{o}dinger-type $f(Q,T)$ model, the cosmological evolution is determined by solving the corresponding set of ordinary differential equations, given by Eqs.~(\ref{87})--(\ref{89}). These equations describe the redshift evolution of the dimensionless Hubble parameter $h(z)$. The numerical solution for $h(z)$ is obtained using the best-fit parameter values derived from the statistical analysis. The Hubble parameter is then expressed as $H(z) = H_0\,h(z)$.

The Hubble residuals are defined as
\begin{equation}
\Delta H(z) = H_{f(Q,T)}(z) - H_{\Lambda \text{CDM}}(z),
\end{equation}
where $H_{f(Q,T)}(z)$ denotes the Hubble parameter predicted by the Schr\"{o}dinger-type $f(Q,T)$ model, and $H_{\Lambda \text{CDM}}(z)$ represents the corresponding $\Lambda$CDM prediction. The residuals $\Delta H(z)$ quantify the deviation of the model from the standard $\Lambda$CDM cosmology.

By examining both the Hubble function $H(z)$ and the residuals $\Delta H(z)$, we evaluate how closely the $f(Q,T)$ model follows the $\Lambda$CDM expansion history and how well it fits the Hubble measurements.

\begin{figure*}[htb]
\begin{subfigure}{.48\textwidth}
\includegraphics[width=\linewidth]{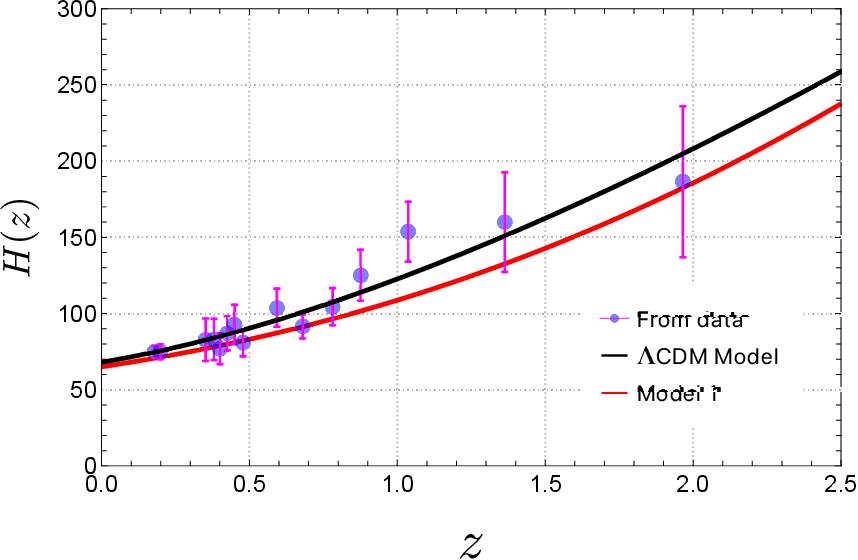}
\end{subfigure}
\hfil
\begin{subfigure}{.47\textwidth}
\includegraphics[width=\linewidth]{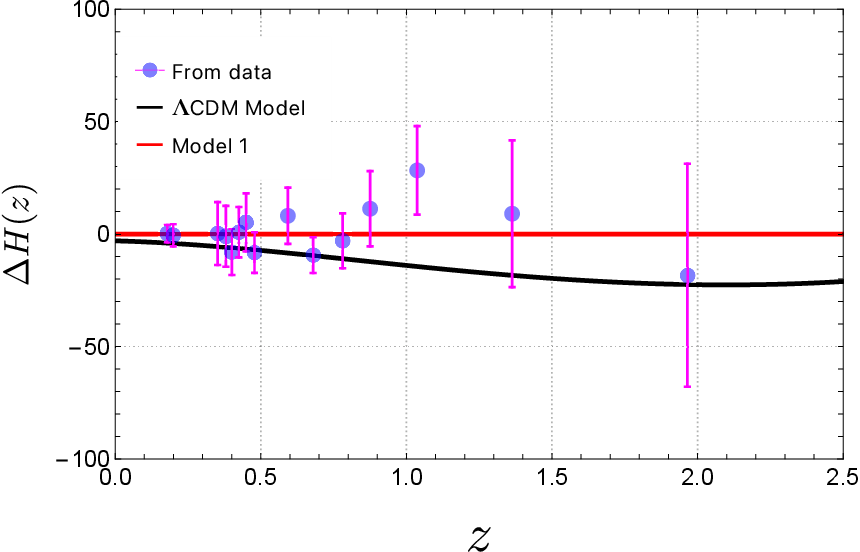}
\end{subfigure}
\caption{This figure shows the comparison of the Schr\"{o}dinger-type $f(Q,T)$ model with the $\Lambda$CDM model using CC measurements. The blue dots show the data points at certain redshifts with their corresponding error bars in magenta. The left panel shows the evolution of the Hubble function, and the right panel shows the Hubble residuals.}\label{fig_13}
\end{figure*}

\subsection{Cosmographic analysis of the Schr\"{o}dinger-type $f(Q,T)$ Model and the $\Lambda$CDM Model}
Cosmography provides a powerful, model-independent approach to explore the kinematical features of the Universe by analyzing the behavior of cosmological observables as functions of redshift. This method does not rely on any specific gravitational theory or assumptions regarding dark energy, but rather on a purely kinematic description of the cosmic expansion. The formalism is based on the Taylor expansion of the scale factor $a(t)$ around the present epoch, allowing the cosmic evolution to be described in terms of measurable quantities such as the Hubble parameter $H(z)$, the deceleration parameter $q(z)$, and higher-order derivatives including the jerk $j(z)$, snap $s(z)$, and jerk $l(z)$ parameters \cite{Visser2004,Cattoen2008,Visser2010,Visser2005,Luongo2011}.

\paragraph{\textbf{Deceleration parameter $q(z)$ and jerk parameter $j(z)$}}
The deceleration parameter, which quantifies the acceleration or deceleration of the cosmic expansion, is defined as
\begin{equation}
q(z) = -1 + (1 + z)\,\frac{H'(z)}{H(z)},
\end{equation}
where $H(z)$ is the Hubble parameter at redshift $z$. A negative value of $q(z)$ indicates an accelerated expansion of the Universe, while a positive value corresponds to a decelerating expansion. Two key cosmographic quantities derived from $q(z)$ are its present-day value $q_0 = q(z=0)$, which characterizes the current expansion rate, and the transition redshift $z_{\text{tr}}$, determined by $q(z_{\text{tr}}) = 0$, representing the epoch at which the Universe transitioned from deceleration to acceleration.

The jerk parameter, which characterizes the rate of change of the acceleration, is given by
\begin{equation}
j(z) = q(z)\,(2\,q(z) + 1) + (1 + z)\,q'(z).
\end{equation}
In the standard $\Lambda$CDM cosmological model, the jerk parameter remains constant at $j(z) = 1$, independent of redshift. Deviations from this value in other cosmological models indicate departures from the $\Lambda$CDM model.
\begin{figure*}[htb]
\begin{subfigure}{.48\textwidth}
\includegraphics[width=\linewidth]{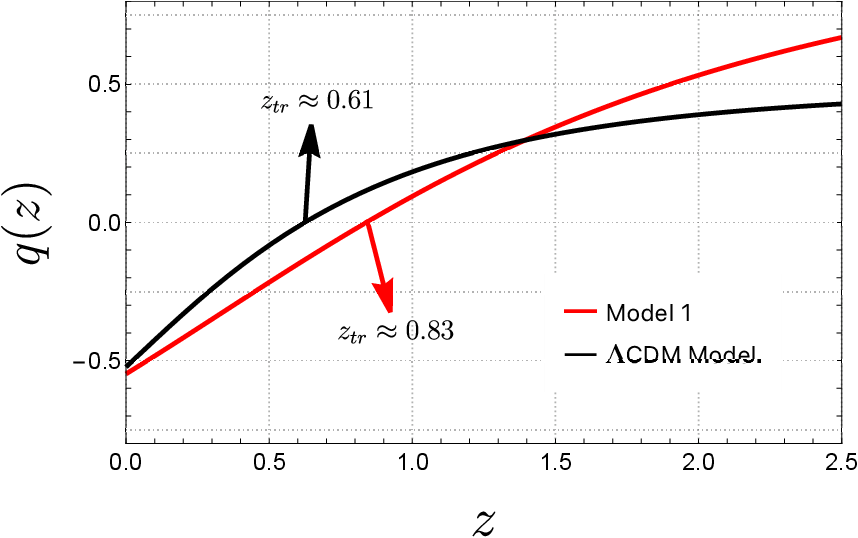}
\end{subfigure}
\hfil
\begin{subfigure}{.48\textwidth}
\includegraphics[width=\linewidth]{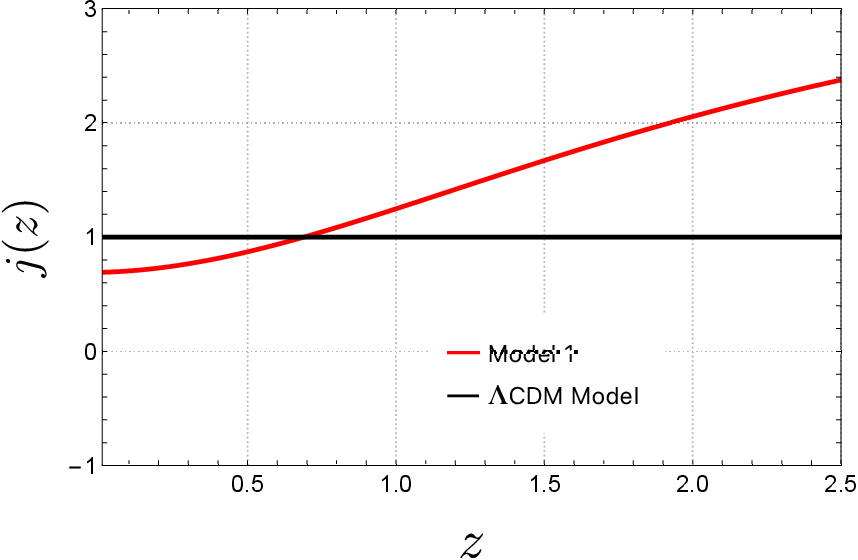}
\end{subfigure}
\caption{The evolution of the cosmographic parameters for the Schr\"{o}dinger-type $f(Q,T)$ model compared with the $\Lambda$CDM model. The left panel shows the deceleration parameter $q(z)$, while the right panel shows the jerk parameter $j(z)$.}\label{fig_14}
\end{figure*}
\paragraph{\textbf{The nonmetricity $\Pi(z)$, the Lagrange multiplier $\Lambda (z)$, and the dimensionless matter density $r(z)$.}}
The nonmetricity function $\Pi (z)$ is the basic indicator of the presence of a Weyl type geometry in the Universe. Its variation with redshift indicates that the nonmetricity did play different roles in different phases of the evolution of the Universe. As for the Lagrange multiplier, it also represents a dynamically evolving quantity, with a significant role in the description of the properties of the late Universe.   

The dimensionless matter density, denoted as $r(z)$ or $\Omega_m(z)$, represents the ratio of the matter energy density to the critical density of the Universe at a given redshift. It quantifies the relative contribution of matter to the total energy budget of the cosmos and evolves with cosmic expansion.
\begin{figure*}[htb]
\begin{subfigure}{.48\textwidth}
\includegraphics[width=\linewidth]{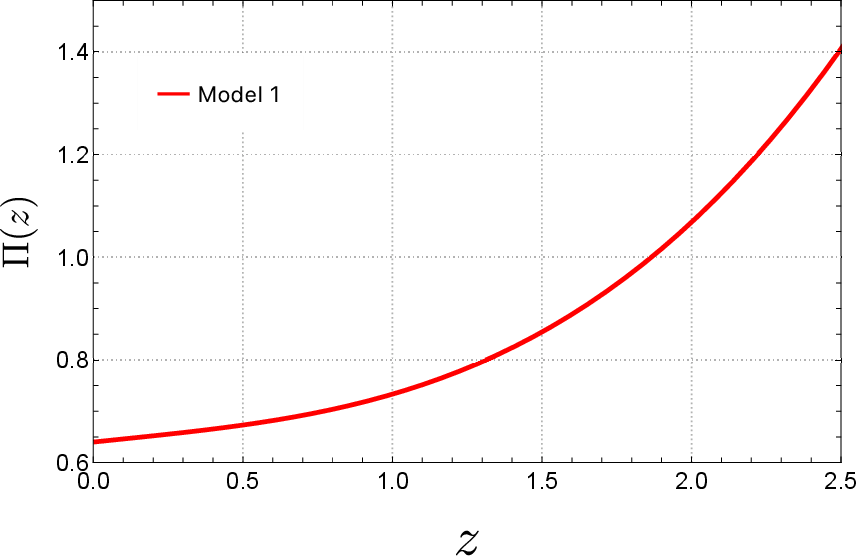}
\end{subfigure}
\hfil
\begin{subfigure}{.48\textwidth}
\includegraphics[width=\linewidth]{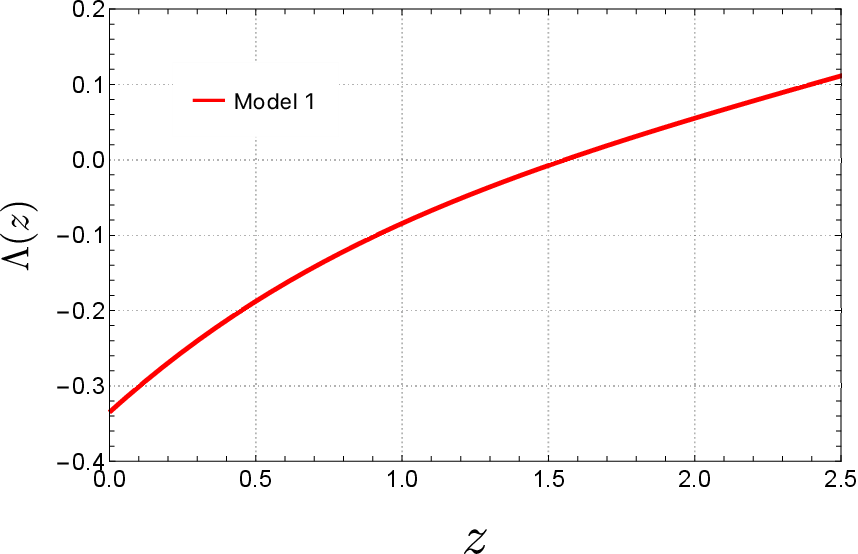}
\end{subfigure}
\caption{The evolution of the dimensionless nonmetricity as a function of redshift (left panel) and the dimensionless Lagrange multiplier (right panel).}\label{fig_15}
\end{figure*}
\begin{figure}
\centering
\includegraphics[scale=0.55]{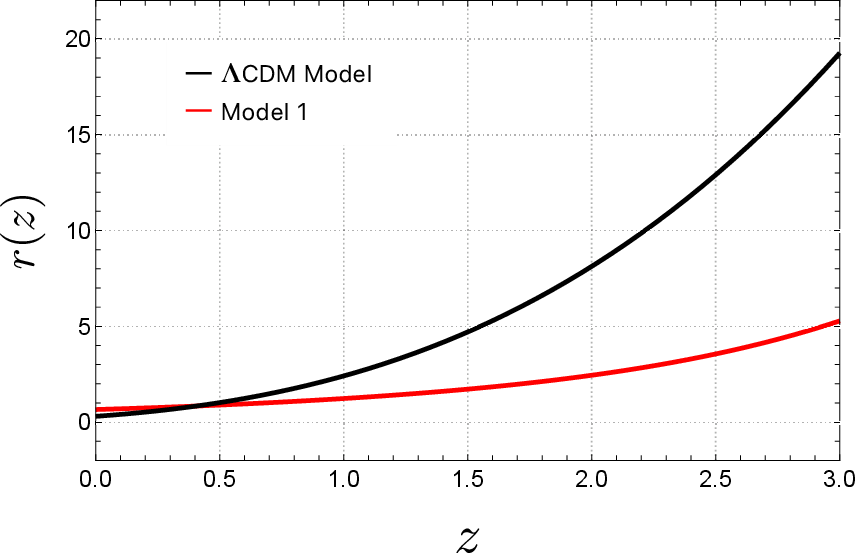}
\caption{The evolution of the dimensionless matter density for the $\Lambda$CDM model and the Schr\"{o}dinger-type $f(Q,T)$ Model with respect to redshift.}\label{fig_16}
\end{figure}

\subsection{Results}

In this section, we provide a detailed description of the results obtained throughout our analysis for the Schr\"{o}dinger-type $f(Q,T)$ model\\

\paragraph{\textbf{MCMC results.}}
In Fig.~\ref{fig_12}, we show the corner plot for the Schr\"{o}dinger-type \( f(Q,T) \) model. The off-diagonal panels of the figure display the 2D marginalized confidence contours at the 68\% and 95\% confidence levels, while the diagonal panels show the 1D marginalized posterior distributions for each parameter of the Schr\"{o}dinger-type \( f(Q,T) \) model. An important feature observed in the results is the negative correlation between the $\beta$ and $\Lambda_0$ parameters. Furthermore, since $r_d$ is treated as a free parameter and determined directly from the data, a negative correlation with $H_0$ is also observed. This correlation is particularly relevant when discussing the $H_0$ tension.

First, we compare the predicted value of $H_0$ obtained from our model. We do not compare it with the Planck~2018 dataset, since the CMB data were not included in our analysis. We only consider late-time datasets, we compare the predicted $H_0$ value from our analysis for the Schr\"{o}dinger-type \( f(Q,T) \) model with that of the $\Lambda$CDM model, so that the comparison remains on an equal footing. The predicted value of the Hubble constant is $H_0 = 67.8 \pm 3.7~\mathrm{km\,s^{-1}\,Mpc^{-1}}$ for the $\Lambda$CDM model and $H_0 = 65.5 \pm 3.8~\mathrm{km\,s^{-1}\,Mpc^{-1}}$ for our model. Compared to $\Lambda$CDM, the Schr\"{o}dinger-type \( f(Q,T) \) model shows a deviation of about $0.43\sigma$ toward a lower value of $H_0$, remaining consistent with $\Lambda$CDM within the $0.5\sigma$ confidence level. The predicted value of the sound horizon is $r_d = 148.5 \pm 7.5~\mathrm{Mpc}$ for the $\Lambda$CDM model and $r_d = 149.5^{+7.5}_{-9.8}~\mathrm{Mpc}$ for the Schr\"{o}dinger-type \( f(Q,T) \) model, indicating only a small deviation of $0.07\sigma$ from the $\Lambda$CDM result. 

The main takeaway from this analysis is that the model does not resolve the $H_0$ tension. This can be understood from two perspectives: to alleviate the $H_0$ tension, one needs to increase the value of $H_0$ while simultaneously decreasing $r_d$, since these two parameters are negatively correlated. However, this behavior is not observed in our case, as the values of both \(H_0\) and \(r_d\) lie within the range of the Planck~2018 results \((H_0 = 67.4 \pm 0.5~\mathrm{km\,s^{-1}\,Mpc^{-1}},~ r_d = 147.09 \pm 0.26~\mathrm{Mpc})\)~\cite{Aghanim2020} and deviate from the local SH0ES determination by Riess~\textit{et~al.} \((H_0 = 73.04 \pm 1.04~\mathrm{km\,s^{-1}\,Mpc^{-1}})\)~\cite{Riess2022}.\\
\paragraph{\textbf{Hubble parameter, and Hubble residual results.}}
In Fig.~\ref{fig_13}, we present a comparison between the $\Lambda$CDM model and the Schr\"{o}dinger-type \( f(Q,T) \) model using the CC measurements, the Hubble function, and its residuals as a function of redshift. The left panel shows the evolution of the Hubble function, where both models closely align with each other at low redshifts ($z < 0.3$). However, at higher redshifts, the Schr\"{o}dinger-type \( f(Q,T) \) model shows a deviation from the $\Lambda$CDM model and predicts slightly smaller values of $H(z)$ compared to $\Lambda$CDM.

The right panel shows the corresponding Hubble residuals, $\Delta H(z)$. A similar behavior can be observed, where the model shows close agreement with the $\Lambda$CDM model at low redshifts and a noticeable deviation at higher redshifts, predicting slightly lower values of $H(z)$. Relative to the data, it can be seen that the residuals for both models lie well within the observational uncertainties, with the Schr\"{o}dinger-type \( f(Q,T) \) model showing a marginally better agreement at intermediate redshifts.\\
\paragraph{\textbf{Cosmographic results.}}
In Fig.~\ref{fig_14}, we show the evolution of the deceleration parameter \( q(z) \) and the jerk parameter \( j(z) \) for the Schr\"{o}dinger-type \( f(Q,T) \) model in comparison with the $\Lambda$CDM model. The left panel shows the redshift evolution of the deceleration parameter. It can be seen that both models exhibit a transition from a decelerated phase (\( q > 0 \)) to an accelerated phase (\( q < 0 \)), but at slightly different transition redshifts. For the $\Lambda$CDM model, the transition occurs at \( z_{\mathrm{tr}} \approx 0.61 \), while for the Schr\"{o}dinger-type \( f(Q,T) \) model, the transition takes place later, at \( z_{\mathrm{tr}} \approx 0.83 \). The present values of the deceleration parameter are \( q_0 = -0.543 \) for Schr\"{o}dinger-type \( f(Q,T) \) model and \( q_0 = -0.519 \) for the $\Lambda$CDM model.

The right panel shows the evolution of the jerk parameter \( j(z) \). For the $\Lambda$CDM model, \( j(z) = 1 \) remains constant at all redshifts. In contrast, the Schr\"{o}dinger-type \( f(Q,T) \) model shows a clear redshift dependence, with \( j(z) < 1 \) at low redshifts and gradually increasing to \( j(z) > 1 \) at higher redshifts.\\
\paragraph{\textbf{Dimensionless nonmetricity, Lagrange multiplier, and matter density results.}}
The dimensionless nonmetricity contribution, $\Pi(z)$, remains positive throughout cosmic evolution, as shown in Fig.~\ref{fig_15}. At late times ($z \lesssim 0.3$), $\Pi(z)$ grows slowly, indicating a stable contribution from the nonmetricity term. Beyond this range, however, $\Pi(z)$ increases more rapidly with redshift, reflecting the strengthening effect of the nonmetricity component in the early Universe.

The right panel of Fig.~\ref{fig_15} shows the evolution of the dimensionless Lagrange multiplier, $\Lambda(z)$. $\Lambda(z)$ evolves smoothly from negative values at low redshifts toward zero as $z$ increases, remaining negative throughout the cosmic history. This behavior suggests that the model maintains a consistent and physically viable evolution of the nonmetricity sector over time.

Fig.~\ref{fig_16} shows the evolution of the dimensionless matter density, \( r(z) \), for the Schr\"{o}dinger-type \( f(Q,T) \) model and the $\Lambda$CDM model as a function of redshift. At low redshifts ($z \lesssim 0.5$), both models exhibit similar behavior, showing that the matter density grows slowly with redshift. However, as the redshift increases, a clear deviation appears between the two models. The Schr\"{o}dinger-type \( f(Q,T) \) model predicts a slower growth rate of \( r(z) \) compared to the $\Lambda$CDM model, resulting in smaller values of the matter density at higher redshifts ($z > 1$).\\
\paragraph{\textbf{Statistical results.}}
Taking the $\Lambda$CDM model as the baseline, we compare its statistical performance with that of the Schr\"{o}dinger-type $f(Q,T)$ model. From Table~\ref{tab_1}, the minimum chi-square values are $\chi^2_{\mathrm{min}} = 1574.88$ for the $\Lambda$CDM model and $\chi^2_{\mathrm{min}} = 1535.62$ for the Schr\"{o}dinger-type $f(Q,T)$ model, yielding $\Delta \chi^2 = -39.26$. The negative value of $\Delta \chi^2$ indicates that Schr\"{o}dinger-type $f(Q,T)$ model provides a better overall fit to the observational data. Similarly, the reduced chi-square values, $\chi^2_{\mathrm{red}} = 0.975$ for $\Lambda$CDM and $\chi^2_{\mathrm{red}} = 0.953$ for the Schr\"{o}dinger-type $f(Q,T)$ model, yields a marginally better agreement with the observational data.

When comparing information criteria, the AIC values are $\text{AIC}_{\Lambda\mathrm{CDM} \text{ Model}} = 1580.88$ and $\text{AIC}_{\text{Schr\"{o}dinger-type} f(Q,T) \text{ Model}} = 1549.62$
, giving a difference of $\Delta \text{AIC} = -31.26$. The BIC values are $\text{BIC}_{\Lambda\mathrm{CDM} \text{ Model}} = 1597.04$ and $\text{AIC}_{\text{Schr\"{o}dinger-type} f(Q,T) \text{ Model}} = 1587.34$, corresponding to $\Delta \text{BIC} = -9.70$. Since both $\Delta \text{AIC}$ and $\Delta \text{BIC}$ are negative, this clearly favors the Schr\"{o}dinger-type $f(Q,T)$ model over $\Lambda$CDM. In particular, the large magnitude of $\Delta \text{AIC}$ ($>10$) and $\Delta \text{BIC}$ ($>5$) indicates strong to decisive statistical evidence in support of the $f(Q,T)$ model.    

The $P$-values also support this result, with $P = 0.758$ for $\Lambda$CDM and $P = 0.909$ for the Schr\"{o}dinger-type $f(Q,T)$ model, indicating that the latter provides a better fit to the data. Finally, the Bayesian evidence difference, $|\Delta \ln \mathcal{Z}_{\Lambda\mathrm{CDM}, f(Q,T)}| = 8.13$, lies within the range of \textit{decisive evidence} on the revised Jeffreys scale, confirming that the Schr\"{o}dinger-type $f(Q,T)$ model is strongly favored over the baseline $\Lambda$CDM cosmology.

\section{Discussions and final remarks}\label{sect5}


In this  work, we have introduced a specific representation of the symmetric teleparallel gravity, in which the dynamics of the gravitational field are described by a the non-metricity scalar, $Q$. We considered a class of theories analogous to the ones introduced in \cite{Q5,Q6}, in which $Q$ is non-minimally coupled to the trace of the matter energy-momentum tensor, $T$. This class of $f(Q,T)$ theories was inspired by $f(R,T)$ theories (see \cite{fRT,book}). In these previous approaches, a Weyl type non-metricity was employed. In the present approach, an important and novel element was that we used a Schr\"odinger type non-metricity instead of a Weyl type. Using a Schr\"{o}dinger non-metricity ensured that the length of a vector does not change under parallel transport. We further restricted the non-metricity to be vectorial, i.e., it was completely determined by a vector field, $\pi_\mu$, dubbed as the Schr\"odinger vector.  The non-metricity scalar was then determined by the norm of the Schr\"{o}dinger vector.

To construct a Schr\"{o}dinger type $f(Q,T)$ gravity theory, we proposed an action in which, the gravitational sector contained three components -- an arbitrary function, $f$, of the non-metricity scalar and the trace of the matter energy-momentum tensor $T$; the strength, $\Pi_{\mu \nu}$, of the Schr\"{o}dinger vector; and the mass of the Schr\"{o}dinger vector, $m$. Further, we imposed the teleparallel condition via a Lagrange multiplier. We obtained the gravitational field equations by varying the action with respect to the metric, the Schr\"{o}dinger vector, and the Lagrange multiplier, respectively. We found that the Schr\"{o}dinger vector satisfies a generalized Proca type equation, which was closely related to the presence of a Lagrange multiplier.
 Note that we have investigated the gravitational action in the framework of the metric formalism. 
 
Using these field equations, we proceeded to investigate the implications of our theory for cosmology. As a first step in this study we have derived the set of generalized Friedmann equations for a flat FLRW type geometry. These equations generalize the Friedmann equations of standard general relativity through the addition of new terms, depending on the Weyl vector and on the Lagrange multiplier. These terms can be interpreted as describing an effective dark energy, of geometric origin. The set of the cosmological evolution equations is enlarged due to the presence of two new evolution equations for the Weyl vector, and for the Lagrange multiplier, respectively. In this respect the Schr\"{o}dinger type $f(Q,T)$  models are more complex mathematically than the standard general relativistic models, but also allow for a much more complicated evolutionary dynamics.

As an application of the obtained formalism we have performed an investigation of the cosmological evolution in both the very early Universe, as well as in the late Universe. For the case of the early Universe we have investigated the possibility of constructing a warm inflationary scenario, in which the production of radiation is determined by geometry. We have thus considered several cases of cosmological evolution, corresponding to particular choices of the functional form of the Weyl vector and of the Lagrange multiplier. In the case of the constant Weyl vector the field equations can be solved exactly. The solution describes an accelerating Universe in which matter in the form of radiation is created during the accelerated expansion of the Universe. However, even though the expansion is not of de Sitter type in the early stages, in the large time limit the expansion becomes exponential, with the Hubble function tending towards a constant value. Hence, the present model may describe both the very early and late stages of cosmological evolution.

A second case which we have considered corresponds to the choice of a constant Lagrange multiplier. In this approach both the effective density and pressure identically vanish, and the evolution reduces to the standard general relativistic one. However, even the cosmological dynamics is general relativistic, the geometric structure of the space-time manifold is not Riemannian, due to the presence of a non-vanishing Weyl vector. But in order to describe warm inflation the presence of a cosmological scalar field is necessary.

We have also considered the full solution of the generalized Friedmann equations for the case $f(Q)=\alpha Q$ in the presence of a time varying Weyl vector, and Lagrange multiplier. The numerical solution indicates that the evolution of the Universe begins from a state in which no matter is present, with the matter being created during the early phases of evolution. The matter density reaches a maximum value at a finite value of the dimensionless time $\tau$, after which it begins to decrease, indicating that the rate of matter production is smaller than the rate of the cosmological expansion. It is interesting to note that in this model the Hubble function is a monotonically increasing function of time.

We have also investigated the possibility of the Schr\"{o}dinger type $f(Q,T)$ models to describe the late stages of evolution of the Universe. We have first formulated the theoretical model of the Schr\"{o}dinger type $f(Q,T)$ gravity model, by adopting the simplest possible form of the Lagrange density as the simple sum of the nonmetricity and trace of the matter energy momentum tensor, assumed to be in the form of dust. Then our main goals were twofold.  Firstly,  we confronted the model with the observations, and, secondly, we obtained the optimal values of the free parameters of the simple Schr\"{o}dinger type $f(Q,T)$ gravity model. 

The Schr\"{o}dinger type $f(Q,T)$ gravity model was compared against several observational datasets, represented by Cosmic Chronometers, Type Ia Supernovae, and Baryon Acoustic Oscillations, respectively. The statistical analysis was performed by using Markov Chain Monte Carlo (MCMC) methods. Finally, the results were also compared with the predictions of the standard $\Lambda$CDM model. 

From the comparison with the observational data, it follows that the cosmological model based on the Schr\"{o}dinger type $f(Q,T)$ gravity is slightly favored over the $\Lambda$CDM model by the cosmological data, or it gives an almost equivalent description of the observational data. On the other hand, the Lagrange multiplier has an interesting variation with respect to the redshift, indicating an evolution from negative values at small redshifts of the order of $z<1.5$ to positive values for $z>1.5$. On the other hand, the Weyl vector is positive and increasing with the redshift (decreasing in time) for all values of $z$.  We would also like to point out that even small deviations in the initial conditions of the Lagrange multiplier and Weyl vector  from the optimal values can lead to significant differences with respect to the observations. 

To conclude the present study, our results indicate that the simple cosmological models based on the Schr\"{o}dinger type $f(Q,T)$ gravity provide a satisfactory description of the observational data obtained from the late Universe, suggesting they represent a viable alternative to the standard cosmological models based on standard general relativity, and Riemannian geometry. Moreover, these models have also the potential for describing the very early stages of cosmological evolution, providing a theoretical framework for a geometric type warm inflation theory.


\section*{Acknowledgments}

We would like to thank Mr. Lehel Csillag and Dr. Anish Agashe for important contribution and significant help in the theoretical formulation and elaboration of this work.  

\begin{appendix}

    \section{Geometric identities with vectorial non-metricity} \label{appendixA}
    In this Appendix, we present some useful geometric identities with vectorial non-metricity, then specialize to the Schrödinger case.
    
    Let us start by proving that the formula \eqref{vectorialdistortion} can be obtained from \eqref{vectorialnon-metricity}. Assuming a vectorial non-metricity of the form \eqref{vectorialnon-metricity}, we compute the required permutations
    \begin{equation}
        -Q_{\lambda \nu \rho}=-b_2 \pi_{\lambda} g_{\nu \rho}-\frac{b_2-2b_1}{2} \left(\pi_\rho g_{\lambda \nu} + \pi_{\nu} g_{\rho \lambda} \right),
    \end{equation}
    \begin{equation}
        Q_{\rho \lambda \nu}=b_2 \pi_\rho g_{\lambda \nu} +\frac{b_2-2b_1}{2}\left(\pi_{\nu} g_{\rho \lambda}+ \pi_{\lambda} g_{\rho \nu} \right),
    \end{equation}
    \begin{equation}
        Q_{\nu \rho \lambda}=b_2 \pi_\nu g_{\rho \lambda}+\frac{b_2-2b_1}{2} \left(\pi_\lambda g_{\nu \rho}+ \pi_{\rho} g_{\nu \lambda} \right).
    \end{equation}
    Hence, the distortion tensor $\tensor{N}{^\mu _\nu _\rho}$ following the conventions of \cite{Csillag2024b} is given by
    \begin{equation}
    \begin{aligned}
        \tensor{N}{^\mu _\nu _\rho}&=\frac{1}{2}g^{\mu \lambda} \left( -Q_{\lambda \nu \rho} + Q_{\rho \lambda \nu} + Q_{\nu \rho \lambda} \right)\\
        &=\frac{1}{2}g^{\mu \lambda} \left(\pi_{\lambda} g_{\nu \rho}\left(-b_2 + \frac{b_2}{2}-b_1 +\frac{b_2}{2} -b_1 \right) \right)\\
        &+\frac{1}{2}g^{\mu \lambda} \left(\pi_{\rho} g_{\lambda \nu} \left( -\frac{b_2}{2} +b_1 + b_2 +\frac{b_2}{2} - b_1\right) \right)\\
        &+\frac{1}{2}g^{\mu \lambda} \left( \pi_{\nu} g_{\rho \lambda} \left(\frac{b_2}{2} - b_1 - \frac{b_2}{2} + b_1 +b_2 \right) \right)\\
        &=-b_1 \pi^\mu g_{\nu \rho}+ \frac{b_2}{2}\delta^{\mu}_{\nu} \pi_{\rho} + \frac{b_2}{2} \delta^{\mu}_{\rho} \pi_{\nu},
    \end{aligned}
    \end{equation}
    which proves \eqref{vectorialdistortion}. For the expression of  $Q_{(\mu \nu \rho)}$, we have 
    \begin{equation}
    \begin{aligned}
        Q_{(\mu \nu \rho)}&=Q_{\mu \nu \rho}+Q_{\rho \mu \nu}+ Q_{\nu \rho \mu}\\
        &= b_2 \pi_{\mu} g_{\nu \rho}+\frac{b_2-2b_1}{2} \left( \pi_{\rho} g_{\mu \nu} + \pi_{\nu} g_{\mu \rho} \right)\\
        &+b_2 \pi_{\rho} g_{\mu \nu}+ \frac{b_2-2b_1}{2} \left(\pi_{\nu} g_{\rho \mu} + \pi_{\mu} g_{\rho \nu} \right)\\
        &+b_2 \pi_{\nu} g_{\rho \mu} +\frac{b_2-2b_1}{2}  \left(\pi_{\mu} g_{\nu \rho} + \pi_{\rho} g_{\nu \mu} \right)\\
        &=\pi_{\mu} g_{\nu \rho} \left(b_2 + \frac{b_2}{2} - b_1 + \frac{b_2}{2} - b_1 \right)\\
        &+\pi_{\rho} g_{\mu \nu} \left( \frac{b_2}{2} - b_1 + b_2 +\frac{b_2}{2} - b_1\right)\\
        &+\pi_{\nu} g_{\mu \rho} \left(\frac{b_2}{2} - b_1 + b_2 + \frac{b_2}{2} - b_1 \right)=0,
    \end{aligned}
    \end{equation}
    which for a non-vanishing vector field $\pi_{\mu}$ translates to the relation  $b_1=b_2$ between the coefficients. Hence, the simplest form of a non-metricity admitting fixed length vectors is given by the choice $b_2=-1$, also called the Yano-Schrödinger connection:
    \begin{equation}
    Q_{\lambda \nu \rho}=-\pi_{\lambda} g_{\nu \rho}+\frac{1}{2} \pi_{\rho} g_{\lambda \nu} + \frac{1}{2}\pi_{\nu} g_{\rho \lambda}.
\end{equation}

Straightforward computations give the following expressions for the non-metricity traces
\begin{equation}
\begin{aligned}
    Q_{\lambda}&=g^{\nu \rho} Q_{\lambda \nu \rho}=-4 \pi_{\lambda}+\frac{1}{2} \pi_{\lambda}+\frac{1}{2} \pi_{\lambda}=-3 \pi_{\lambda}, \\
    \widetilde{Q}_{\nu}&=g^{\lambda \rho} Q_{\lambda \nu \rho}=- \pi_{\nu}+\frac{1}{2} \pi_{\nu} +2 \pi_{\nu}=\frac{3}{2} \pi_{\nu}.
\end{aligned}
\end{equation}

Let us note that the traces satisfy the following identity
\begin{equation}
    Q_{\lambda}=-2 \widetilde{Q}_{\lambda}.
\end{equation}

The superpotential can be obtained in a relatively straightforward way, with a long algebraic computation, which reads
\begin{eqnarray}
    P^{\alpha \mu \nu}&=&-\frac{1}{4} \left({ - \pi^\alpha g^{\mu \nu}} {+ \frac{1}{2} \pi^{\nu} g^{\alpha \mu}} {+ \frac{1}{2} \pi^\mu g^{\nu \alpha}}\right)\nonumber\\
    &&+\frac{1}{4} \left({- \pi^{\mu} g^{\nu \alpha}}  {+ \frac{1}{2} \pi^{\alpha} g^{\mu \nu}} {+ \frac{1}{2} \pi^{\nu} g^{\mu \alpha}}  \right)\nonumber\\
    &&+ \frac{1}{4} \left({- \pi^{\nu} g^{\mu \alpha}} {+\frac{1}{2} \pi^{\alpha} g^{\mu \nu}}  {+\frac{1}{2} \pi^{\mu} g^{\nu \alpha}} \right)\nonumber\\
  &&{+\frac{1}{4} \left(-3 \pi^{\alpha} - \frac{3}{2} \pi^{\alpha} \right) g^{\mu \nu}}
    {+\frac{3}{8} g^{\alpha \mu} \pi^{\nu}} {+ \frac{3}{8} g^{\alpha \nu} \pi^{\mu}}\nonumber\\
    &=& \pi^{\alpha} g^{\mu \nu} \left( {\frac{1}{4} + \frac{1}{8} + \frac{1}{8}- \frac{9}{8}} \right)\nonumber\\
    &&+ \pi^{\nu} g^{\alpha \mu} \left({-\frac{1}{8} + \frac{1}{8} -\frac{1}{4} + \frac{3}{8}} \right)\nonumber\\
    &&+\pi^{\mu} g^{\nu \alpha} \left( {-\frac{1}{8} -\frac{1}{4} + \frac{1}{8} + \frac{3}{8}}\right)\nonumber\\
    &=& -\frac{5}{8}\pi^{\alpha} g^{\mu \nu} + \frac{1}{8} \pi^{\nu} g^{\alpha \mu} + \frac{1}{8} \pi^{\mu} g^{\nu \alpha}.
\end{eqnarray}

Substituting this expression into the non-metricity scalar, we immediately obtain
\bea
    Q&=&\left( \pi_{\alpha} g_{\mu \nu} - \frac{1}{2} \pi_{\nu} g_{\alpha \mu} - \frac{1}{2} \pi_{\mu} g_{\alpha \nu} \right) \nonumber\\
    &&\times \left( -\frac{5}{8}\pi^{\alpha} g^{\mu \nu} + \frac{1}{8} \pi^{\nu} g^{\alpha \mu} + \frac{1}{8} \pi^{\mu} g^{\nu \alpha}\right)\nonumber\\
    &=&-\frac{5}{2} \pi_\alpha \pi^\alpha + \frac{1}{8} \pi^\mu \pi_\mu +\frac{1}{8} \pi^\nu \pi_\nu +\frac{5}{16} \pi^\mu \pi_\mu \nonumber\\
    &&- \frac{1}{4} \pi_\nu \pi^\nu - \frac{1}{16} \pi^\alpha \pi_\alpha
    +\frac{5}{16} \pi^\nu \pi_\nu - \frac{1}{16} \pi_\alpha \pi^\alpha -\frac{1}{4} \pi_\mu \pi^\mu \nonumber\\
    &=&-\frac{9}{4} \pi^{\alpha} \pi_{\alpha}.
\eea
\section{Derivation of the field equations}\label{appendixB}

\subsection{Metric field equations}

One has to vary the action with respect the three independent variables $\{g_{\mu \nu}, \pi_{\mu}, \lambda_{\mu}\}$. We first compute the variations with respect to the metric, and introduce the notations
\begin{equation}
\begin{aligned}
    S_1&:=\kappa^2 \int f(Q,T) \sqrt{-g} \mathrm{d}^4 x,\\
    S_2&:=-\frac{1}{2} \int \left[\frac{1}{2} \Pi_{\mu \nu} \Pi^{\mu \nu}+m^2 \pi_\mu \pi^\mu \right] \sqrt{-g} \mathrm{d}^4x,\\
    S_3&:= \int \mathcal{L}_{m} \sqrt{-g} \mathrm{d}^4x,\\
    S_4&:= \int \left \{ \lambda \left(\overset{\circ}{R} + \frac{9}{2} \overset{\circ}{\nabla}_{\mu} \pi^{\mu} - \frac{9}{4} \pi_\mu \pi^\mu \right) \right \} \sqrt{-g} \mathrm{d}^4x.
\end{aligned}
\end{equation}
In these notations, the field equations are obtained as
\begin{equation}
    \frac{ \delta S}{\delta g^{\mu \nu}}=\frac{\delta S_1}{\delta g^{\mu \nu}}+\frac{\delta S_2}{\delta g^{\mu \nu}} +\frac{\delta S_3}{\delta g^{\mu \nu}}+ \frac{\delta S_4}{\delta g^{\mu \nu}}=0.
\end{equation}

Each term is calculated separately. The first term evaluates to
\begin{equation}\label{variationS1}
\begin{split}
    \frac{\delta S_1}{\delta g^{\mu \nu}} = \kappa^2 \int \Bigg\{
    & \frac{\partial f(Q,T)}{\partial Q} \frac{\delta Q}{\delta g^{\mu \nu}}
    + \frac{\partial f(Q,T)}{\partial T} \frac{\delta T}{\delta g^{\mu \nu}} \\
    & - \frac{1}{2} g_{\mu \nu} f(Q,T)
    \Bigg\} \sqrt{-g}\, \mathrm{d}^4x ,
\end{split}
\end{equation}
where we used the identity
\begin{equation}\label{variationsqrt}
    \delta \sqrt{-g}=-\frac{\sqrt{-g}}{2} g_{\mu \nu} \delta g^{\mu \nu} \iff \frac{\delta{\sqrt{-g}}}{\delta g^{\mu \nu}}=-\frac{1}{2} g_{\mu \nu} \sqrt{-g}. 
\end{equation}

Using the explicit form of the non-metricity scalar \eqref{non-metricityscalarschrexpression}, we calculate
\begin{equation}\label{variationq}
    \frac{\delta Q}{\delta g^{\mu \nu}}=\frac{\delta \left(-\frac{9}{4} \pi_\alpha \pi_\beta g^{\alpha \beta} \right)}{\delta g^{\mu \nu}}= -\frac{9}{4}\pi_{\alpha }\pi_{\beta} \delta^{\alpha}_{\mu} \delta^{\beta}_{\nu}= -\frac{9}{4} \pi_{\mu} \pi_{\nu}.
\end{equation}

For the variation of the trace of the energy–momentum tensor, we obtain
\begin{equation}\label{variationt}
    \frac{\delta T}{\delta g^{\mu \nu}}=\frac{\delta \left( T_{\alpha \beta} g^{\alpha \beta} \right)}{\delta g^{\mu \nu}}=-T_{\mu \nu} + g_{\mu \nu} L_{m}.
\end{equation}

Substituting equations \eqref{variationsqrt}-\eqref{variationt} into \eqref{variationS1} leads to
\begin{equation}\label{rewrite}
\begin{split}
    \frac{\delta S_1}{\delta g^{\mu \nu}} = \kappa^2 \int \Bigg\{
    & -\frac{9}{4} \frac{\partial f(Q,T)}{\partial Q} \pi_{\mu} \pi_{\nu}\\ 
    & + \frac{\partial f(Q,T)}{\partial T} \left( - T_{\mu \nu} + g_{\mu \nu} L_{m} \right) \\
    & - \frac{1}{2} g_{\mu \nu} f(Q,T)
    \Bigg\} \sqrt{-g}\, \mathrm{d}^4x .
\end{split}
\end{equation}

Introducing the notations
\begin{equation}
   f(Q,T):=f, \; \; \frac{\partial f(Q,T)}{\partial Q}:=f_{Q}, \; \; \frac{\partial f(Q,T)}{\partial T}:= f_{T},
\end{equation}
we can rewrite equation \eqref{rewrite} as
\begin{equation}
\begin{split}
    \frac{\delta S_1}{\delta g^{\mu \nu}} = \kappa^2 \int \Bigg\{
    & -\frac{9}{4} f_Q\, \pi_{\mu} \pi_{\nu}
    + f_T \left( - T_{\mu \nu} + g_{\mu \nu} L_{m} \right) \\
    & - \frac{1}{2} g_{\mu \nu} f
    \Bigg\} \sqrt{-g}\, \mathrm{d}^4x .
\end{split}
\end{equation}

To compute $S_2$, we use the identity

\begin{equation}
\begin{split}
& \frac{\delta}{\delta g^{\mu \nu}}
\left(g^{\alpha \gamma} g^{\beta \delta} \Pi_{\alpha \beta} \Pi_{\gamma \delta}\right) \\
&= \delta^{\alpha}_{\mu} \delta^{\gamma}_{\nu} g^{\beta \delta} \Pi_{\alpha \beta} \Pi_{\gamma \delta}
+ g^{\alpha \gamma} \delta^{\beta}_{\mu} \delta^{\delta}_{\nu} \Pi_{\alpha \beta} \Pi_{\gamma \delta} \\
&= g^{\beta \delta} \Pi_{\mu \beta} \Pi_{\nu \delta}
+ g^{\alpha \gamma} \Pi_{\alpha \mu} \Pi_{\gamma \nu}\\
&= 2 \Pi^{\sigma}_{\mu} \Pi_{\nu \sigma}.
\end{split}
\end{equation}

Consequently, for $S_2$ we have
\begin{equation}
\begin{split}
    \frac{ \delta S_2}{\delta g^{\mu \nu}} &= \\
    & -\frac{1}{4} \int \left \{ \frac{ \delta \left( g^{\alpha \gamma} g^{\beta \delta} \Pi_{\alpha \beta} \Pi_{\gamma \delta} \right)}{\delta g^{\mu \nu}} \sqrt{-g} + \frac{\delta \sqrt{-g}}{\delta g^{\mu \nu}} \Pi_{\rho \lambda} \Pi^{\rho \lambda}  \right \} \mathrm{d}^4x \\
    &-\frac{1}{2} \int \left\{ \frac{ \delta  \left(m^2 \pi_{\lambda} \pi_{\rho} g^{\lambda \rho} \right)}{\delta g^{\mu \nu}}\sqrt{-g} + m^2 \pi_\rho \pi^\rho \frac{ \delta \sqrt{-g}}{\delta g^{\mu \nu}} \right\} \mathrm{d}^4 x \\
    &= \int \Bigg\{
    -\frac{1}{2} \Pi_{\mu \beta} {\Pi}{_\nu ^\beta}
    + \frac{1}{8} \Pi_{\rho \lambda} \Pi^{\rho \lambda} g_{\mu \nu}
    - \frac{1}{2} m^2 \pi_\mu \pi_\nu \\[4pt]
    & \quad
    + \frac{1}{4} m^2 g_{\mu \nu} \pi_{\rho} \pi^{\rho}
    \Bigg\} \sqrt{-g}\, \mathrm{d}^4x .
\end{split}
\end{equation}

The variation  $S_3$ follows directly from the definition of the energy-momentum tensor
\begin{equation}
   T_{\mu \nu}=-\frac{2}{\sqrt{-g}} \frac{ \delta \left( \sqrt{-g} \mathcal{L}_m \right)}{\delta g^{\mu \nu}},
\end{equation}
which directly implies
\begin{equation}
     \frac{\delta S_3}{\delta g^{\mu \nu}}= \int \frac{ \delta \left( \mathcal{L}_{m} \sqrt{-g} \right)}{\delta g^{\mu \nu}} \mathrm{d^4} x=-\frac{1}{2}\int T_{\mu \nu} \sqrt{-g} \mathrm{d}^4 x.
\end{equation}

We now compute the variation of $S_4$ with respect to the metric
\begin{equation}
\begin{aligned}
  \delta{S}_4= \int &\mathrm{d}^4 x \sqrt{-g} \left( \delta \overset{\circ}{R}+ \frac{9}{2}  \delta \left(\overset{\circ}{\nabla}_{\mu} \pi^{\mu} \right) - \frac{9}{4} \delta \left( \pi_{\mu} \pi_{\nu} g^{\mu \nu}\right) \right) \lambda
  \\
  =\int &\mathrm{d}^4x \sqrt{-g} \Bigg(\lambda \overset{\circ}{R}_{\mu \nu}+ g_{\mu \nu} \overset{\circ}{\Box}\lambda -\overset{\circ}{\nabla}_{\mu} \overset{\circ}{\nabla}_{\nu} \lambda -\frac{9}{4} \pi_{\mu} \pi_{\nu} \lambda
  \\
  &-\frac{9}{4} \pi_\mu \overset{\circ}{\nabla}_{\nu} \lambda - \frac{9}{4} \pi_\nu \overset{\circ}{\nabla}_{\mu} \lambda + \frac{9}{4} g_{\mu \nu} \overset{\circ}{\nabla}_{\rho}\left( \lambda \pi^\rho \right)\Bigg) \delta g^{\mu \nu}.
\end{aligned}
\end{equation}
Putting these results together, one obtains
\begin{equation*}
\begin{split}
& -\frac{9}{4} \kappa^2 f_{Q} \pi_{\mu} \pi_{\nu} 
+ \kappa^2 f_{T} \left( -T_{\mu \nu} + g_{\mu \nu} L_{m} \right)
- \frac{1}{2} \kappa^2 g_{\mu \nu} f \\
& - \frac{1}{2} \Pi_{\mu \beta} {\Pi}{_\nu ^\beta}
+ \frac{1}{8} \Pi_{\rho \lambda} \Pi^{\rho \lambda} g_{\mu \nu} - \frac{1}{2} m^2 \pi_\mu \pi_\nu
+ \frac{1}{4} m^2 g_{\mu \nu} \pi_{\rho} \pi^{\rho}\\
& - \frac{1}{2} T_{\mu \nu}
+ \lambda \overset{\circ}{R}_{\mu \nu}
+ g_{\mu \nu} \overset{\circ}{\Box} \lambda  - \overset{\circ}{\nabla}_{\mu} \overset{\circ}{\nabla}_{\nu} \lambda
- \frac{9}{4} \pi_{\mu} \pi_{\nu} \lambda
- \frac{9}{4} \pi_\mu \overset{\circ}{\nabla}_{\nu} \lambda \\
& - \frac{9}{4} \pi_\nu \overset{\circ}{\nabla}_{\mu} \lambda  + \frac{9}{4} g_{\mu \nu} 
  \overset{\circ}{\nabla}_{\rho} \left( \lambda \pi^\rho \right)
= 0 .
\end{split}
\end{equation*}

Rearranging this equation immediately leads to 
\begin{equation}\label{pluginhere}
\begin{split}
& \frac{1}{2} T_{\mu \nu}
+ \frac{1}{2} \Pi_{\mu \beta}{\Pi}{_\nu ^\beta}
- \frac{1}{8} \Pi_{\rho \lambda} \Pi^{\rho \lambda} g_{\mu \nu}
+ \frac{1}{2} m^2 \pi_{\mu} \pi_{\nu}\\
& - \frac{1}{4} m^2 g_{\mu \nu} \pi_\rho \pi^\rho
- \kappa^2 f_{T} \left( - T_{\mu \nu} + g_{\mu \nu} L_{m} \right) \\[4pt]
& = -\frac{9}{4} \kappa^2 f_{Q} \pi_{\mu} \pi_{\nu}
- \frac{1}{2} g_{\mu \nu}
- \frac{1}{2} \kappa^2 g_{\mu \nu} f \\
& + \lambda \left( \overset{\circ}{R}_{\mu \nu}
- \frac{9}{4} \pi_{\mu} \pi_{\nu}
+ \frac{9}{4} g_{\mu \nu} \overset{\circ}{\nabla}_{\rho} \pi^{\rho} \right) \\[4pt]
& \quad + \frac{9}{4} g_{\mu \nu} \pi^{\rho} \overset{\circ}{\nabla}_{\rho} \lambda
- \frac{9}{4} \pi_{\mu} \overset{\circ}{\nabla}_{\nu} \lambda
- \frac{9}{4} \pi_{\nu} \overset{\circ}{\nabla}_{\mu} \lambda \\[4pt]
& \quad + g_{\mu \nu} \overset{\circ}{\Box} \lambda
- \overset{\circ}{\nabla}_{\mu} \overset{\circ}{\nabla}_{\nu} \lambda .
\end{split}
\end{equation}

Recall that varying with respect to the Lagrange multiplier gives the flat constraint
\begin{equation}
     \overset{\circ}{R}+\frac{9}{2} \overset{\circ}{\nabla}_{\rho} \pi^{\rho}-\frac{9}{4} \pi_{\rho} \pi^{\rho}=0,
\end{equation}
from which expressing the Ricci scalar of the Levi-Civita connection yields
\begin{equation}\label{pluginthis}
    \frac{1}{2} g_{\mu \nu}\overset{\circ}{R}=\frac{9}{8} g_{\mu \nu} \pi_{\rho} \pi^{\rho} -\frac{9}{4} g_{\mu \nu} \overset{\circ}{\nabla}_{\rho} \pi^{\rho}.
\end{equation}

Substituting \eqref{pluginthis} into \eqref{pluginhere} results in the final form of the field equations \eqref{fieldeqn1}
\begin{equation}
\begin{split}
& \frac{1}{2} T_{\mu \nu}
+ \frac{1}{2} \Pi_{\mu \beta} {\Pi}{_\nu ^\beta}
- \frac{1}{8} \Pi_{\rho \lambda} \Pi^{\rho \lambda} g_{\mu \nu}
+ \frac{1}{2} m^2 \pi_{\mu} \pi_{\nu} \\
& - \frac{1}{4} m^2 g_{\mu \nu} \pi_\rho \pi^\rho
- \kappa^2 f_{T} \left( - T_{\mu \nu} + g_{\mu \nu} L_{m} \right) \\[4pt]
& = -\frac{9}{4} \kappa^2 f_{Q} \pi_{\mu} \pi_{\nu}
- \frac{1}{2} g_{\mu \nu}
- \frac{1}{2} \kappa^2 g_{\mu \nu} f
+ \\
& \lambda \left(
     \overset{\circ}{R}_{\mu \nu}
     - \frac{1}{2} g_{\mu \nu} \overset{\circ}{R}
     - \frac{9}{4} \pi_{\mu} \pi_{\nu}
     + \frac{9}{8} g_{\mu \nu} \pi_{\rho} \pi^{\rho}
  \right) \\[4pt]
& \quad + \frac{9}{4} g_{\mu \nu} \pi^{\rho} 
  \overset{\circ}{\nabla}_{\rho} \lambda
- \frac{9}{4} \pi_{\mu} 
  \overset{\circ}{\nabla}_{\nu} \lambda
- \frac{9}{4} \pi_{\nu} 
  \overset{\circ}{\nabla}_{\mu} \lambda \\[4pt]
& \quad + g_{\mu \nu} 
  \overset{\circ}{\Box} \lambda
- \overset{\circ}{\nabla}_{\mu} 
  \overset{\circ}{\nabla}_{\nu} \lambda .
\end{split}
\end{equation}

\subsection{Schrödinger vector field equations}
Since the matter energy-momentum tensor is independent from the Schrödinger vector $\pi_{\mu}$, one has
\begin{equation}
    \frac{\delta f}{\delta \pi_\mu}=f_{Q} \frac{\delta Q}{\delta \pi_{\mu}}=-\frac{9}{2} f_{Q} \pi^{\mu},
\end{equation}
where we used \eqref{non-metricityscalarschrexpression}. It is now straightforward to verify the following results,
\begin{equation}
\frac{\delta\left(\Pi_{\mu \nu} \Pi^{\mu \nu} \right)}{\delta \pi_{\mu}}=4 \left(- \overset{\circ}{\Box} \pi^{\mu}+ \overset{\circ}{\nabla}_{\nu} \overset{\circ}{\nabla} \tensor{}{^\mu} \pi^{\nu} \right), 
\end{equation}
\begin{equation}
    \frac{\delta \left( \pi_{\rho} \pi^\rho \right)}{\delta \pi_{\mu}}=2 \pi^{\mu},
\end{equation}

\begin{equation}
    \lambda \frac{\delta R}{\delta \pi_{\mu}}=\lambda \frac{9}{2} \frac{\delta \left( \overset{\circ}{\nabla}_{\alpha} \pi^{\alpha}\right)}{\delta \pi_{\mu}}-\frac{9}{4}\lambda \frac{ \delta( \pi_{\rho} \pi^{\rho})}{\delta \pi^{\mu}}=-\frac{9}{2} \overset{\circ}{\nabla}_{\mu} \lambda -\frac{9}{2} \lambda \pi_{\mu}.
\end{equation}

Hence, the evolution equation of the Schrödinger non-metricity vector is given by \eqref{vectorfieldeom}
\begin{equation}
    \overset{\circ}{\Box} \pi^{\mu} - \overset{\circ}{\nabla}_{\nu} \overset{\circ}{\nabla}{^\mu} \pi^{\nu}- \left(\frac{9}{2} \kappa^2 f_{Q} +m^2 +\frac{9}{2} \lambda \right) \pi^{\mu}=\frac{9}{2} \overset{\circ}{\nabla}_{\mu} \lambda.
\end{equation}
\section{Derivation of the Friedmann equations}\label{appendixC}
The non-zero components of the Christoffel symbol and the corresponding Ricci tensor
are
\begin{equation}
\begin{aligned} &\overset{\circ}{\Gamma}\tensor{}{^0 _i_i}=a\dot{a}, \quad
\overset{\circ}{\Gamma}\tensor{}{^i_{i0}}=\overset{\circ}{\Gamma}\tensor{}{^i_{0i}}=\frac{\dot{a}}{a}, \\
&\overset{\circ}{R}_{00}=-\frac{3\Ddot{a}}{a}, \quad
\overset{\circ}{R}_{ii}=2\dot{a}^2+a\Ddot{a}, \end{aligned}
\end{equation}
and therefore $\overset{\circ}{R}=6(\dot{a}^2+a\Ddot{a})/a^2$.

To go forward with the field equations, we still need the components of $%
\overset{\circ }{\nabla }_{\mu }\pi _{\nu }$, $\overset{\circ }{\Box }\pi
_{\mu }$, $\overset{\circ }{\nabla }_{\nu }\overset{\circ }{\nabla }_{\mu
}\pi ^{\nu }$ and $\overset{\circ }{\nabla }_{\mu }\overset{\circ }{\nabla }%
_{\nu }\lambda $. The non-zero components of $\overset{\circ }{\nabla }_{\mu
}\pi _{\nu }=\partial _{\mu }\pi _{\nu }-\overset{\circ}{\Gamma}\tensor{}{^\lambda _\mu _\nu}\pi
_{\lambda }$ are
\begin{equation}
\overset{\circ }{\nabla }_{0}\pi _{0}=\dot{\pi},\quad \overset{\circ }{%
\nabla }_{i}\pi _{i}=-a\dot{a}\pi ,
\end{equation}%
thus $\overset{\circ }{\nabla }_{\mu }\pi ^{\mu }=-\dot{\pi}-3\dot{a}\pi /a$%
. Considering that
\begin{equation}
\overset{\circ }{\nabla }_{\rho }\overset{\circ }{\nabla }_{\mu }\pi _{\nu
}=\partial _{\rho }\left(\overset{\circ }{\nabla }_{\mu }\pi _{\nu }\right)-\overset{\circ}{\Gamma}\tensor{}{^\sigma _\rho_\mu}
\overset{\circ }{\nabla }_{\sigma }\pi _{\nu }-\overset{\circ}{\Gamma}\tensor{}{^\sigma _\rho _\nu}\overset{\circ }{\nabla }_{\mu }\pi _{\sigma },
\end{equation}%
we can get
\begin{equation}
\begin{aligned}
\overset{\circ}{\Box}\pi_0&=g^{00}\partial_0 \left(\overset{\circ}{\nabla}_0%
\pi_0 \right)-3g^{ii}\overset{\circ}{\Gamma}\tensor{}{^0_i_i}\overset{\circ}{\nabla}_0\pi_0-3g^{ii}
\overset{\circ}{\Gamma}\tensor{}{^i_i_0}\overset{\circ}{\nabla}_i\pi_i\\
&=-\Ddot{\pi}-\frac{3\dot{a}\dot{\pi}}{a}+\frac{3\dot{a}^2\pi}{a^2},\\
\overset{\circ}{\Box}\pi_i&=g^{ii}\partial_i\left(\overset{\circ}{\nabla}_i%
\pi_i\right)-g^{\rho\nu}\overset{\circ}{\Gamma}\tensor{}{^i _\rho _\nu}\overset{\circ}{\nabla}_i\pi_i-g^{ii} \overset{\circ}{\Gamma} \tensor{}{^\sigma_{ii}}\overset{\circ}{\nabla}_i\pi_\sigma\\
&-g^{00}\overset{\circ}{\Gamma}\tensor{}{^\sigma_{0i}}\overset{\circ}{\nabla}_0\pi_\sigma=0,
\end{aligned}
\end{equation}%
and
\begin{equation}
\begin{aligned}
\overset{\circ}{\nabla}_\nu\overset{\circ}{\nabla}_0\pi^\nu&=g^{00}%
\partial_0\left(\overset{\circ}{\nabla}_0\pi_0\right)-3g^{ii}\overset{\circ}{\Gamma}\tensor{}{^0_{ii}}\overset{
\circ}{\nabla}_0\pi_0-3g^{ii}\overset{\circ}{\Gamma}\tensor{}{^i_{i0}}\overset{\circ}{\nabla}_i\pi_i \\
&=-\Ddot{\pi}-\frac{3\dot{a}\dot{\pi}}{a}+\frac{3\dot{a}^2\pi}{a^2},\\
\overset{\circ}{\nabla}_\nu\overset{\circ}{\nabla}_i\pi^\nu&=g^{ii}%
\partial_i\left(\overset{\circ}{\nabla}_i\pi_i\right)-g^{ii}\overset{\circ}{\Gamma}\tensor{}{^\sigma_{ii}}\overset{%
\circ}{\nabla}_\sigma\pi_i-g^{00}\overset{\circ}{\Gamma}\tensor{}{^\sigma_{0i}}\overset{\circ}{\nabla}_%
\sigma\pi_0\\ &-g^{\rho\nu}\overset{\circ}{\Gamma}\tensor{}{^i_{\rho\nu}}\overset{\circ}{\nabla}_i\pi_i
=0. \end{aligned}
\end{equation}

Finally, using $\overset{\circ }{\nabla }_{\mu }\overset{\circ }{\nabla }_{\nu
}\lambda =\partial _{\mu }\partial _{\nu }\lambda -\overset{\circ}{\Gamma} \tensor{}{^\rho _\mu _\nu}\partial _{\rho }\lambda $ one gets
\begin{equation}
\begin{aligned}
\overset{\circ}{\nabla}_0\overset{\circ}{\nabla}_0\lambda=\Ddot{\lambda}, \; \;
\overset{\circ}{\nabla}_i\overset{\circ}{\nabla}_i\lambda=-a\dot{a}{\dot
\lambda}, \; \; \overset{\circ}{\Box}\lambda=-\Ddot{\lambda}-\frac{3\dot{a}\dot{%
\lambda}}{a}. \end{aligned}
\end{equation}

With the help of the above results, we arrive at the Friedmann equations
\begin{equation}
\begin{aligned}
&\frac{1}{2}\rho+\frac{1}{4}m^2\pi^2+\kappa^2f_T(\rho+p)+\frac{9}{4}%
\kappa^2f_Q\pi^2-\frac{1}{2}\kappa^2f\\
=&\lambda\left(-3\frac{\Ddot{a}}{a}+3 \frac{\Ddot{a}}{a}+3 \frac{\dot
a^2}{a^2} -\frac{9}{4} \pi^2 +\frac{9}{8} \pi^2
\right)-\frac{9}{4}\pi\dot{\lambda}\\
&+\Ddot{\lambda}+\frac{3\dot{a}\dot{\lambda}}{a}-\Ddot{\lambda},
\end{aligned}
\end{equation}
\begin{equation}
\begin{aligned} &\frac{1}{2} p a^2 + \frac{1}{4} m^2 a^2 \pi^2 +\frac{1}{2}
\kappa^2 a^2 f\\ &= \lambda \left(a \ddot a +2 \dot{a}^2 -3\ddot{a} a -3
\dot{a}^2 - \frac{9}{8}a^2 \pi^2 \right)-\frac{9}{4} a^2 \pi \dot \lambda\\
&- a^2 \left( \ddot \lambda + \frac{3 \dot a \dot \lambda}{a} \right) + a
\dot{a} \dot \lambda. \end{aligned}
\end{equation}

The evolution equation of the vector field becomes
\begin{equation}
-\left( m^{2}+\frac{9}{2}\kappa ^{2}f_{Q}+\frac{9}{2}\lambda \right) \pi =%
\frac{9}{2}\dot{\lambda},
\end{equation}
and the constraint $R=0$ can be written as
\begin{equation}
6\left( \frac{\dot{a}^{2}}{a^{2}}+\frac{\Ddot{a}}{a}\right) -\frac{9}{2}%
\left( \dot{\pi}+\frac{3\dot{a}}{a}\pi \right) +\frac{9}{4}\pi ^{2}=0.
\end{equation}

By reordering the terms and introducing the Hubble function $H=\frac{\dot a}{a}$, the Friedmann equations take the equivalent form
\begin{equation}
\begin{aligned} \kappa^2 f_{T}(\rho+p)&+\frac{1}{2} \rho=\frac{\kappa^2}{2}
f- \left( \frac{9}{4} \kappa^2 f_{Q} + \frac{1}{4} m^2 \right) \pi^2\\
&-3\lambda \left(\frac{3}{8} \pi^2 -H^2 \right)-3 \dot \lambda \left(
\frac{3}{4} \pi - H \right), \end{aligned}
\end{equation}%
\begin{eqnarray}
-\frac{1}{2}p &=&\frac{\kappa ^{2}f}{2}+\frac{1}{4}m^{2}\pi ^{2}+\lambda
\left( 3H^{2}+{2\dot{H}}+\frac{9}{8}\pi ^{2}\right)   \notag \\
&&+\dot{\lambda}\left( \frac{9}{4}\pi +2H\right) +\ddot{\lambda}.
\end{eqnarray}

To bring them closer to the standard general relativistic form, we express
them as
\begin{eqnarray}
3H^{2} &=&\frac{\rho }{2\lambda }+\frac{1}{\lambda }\Bigg(\kappa
^{2}f_{T}(\rho +p)-\frac{\kappa ^{2}f}{2}+\frac{9}{4}\kappa ^{2}f_{Q}{\pi
^{2}}  \notag \\
&&+\frac{1}{4}m^{2}\pi ^{2}+\frac{9}{8}{\lambda }\pi ^{2}+\frac{9}{4}\dot{%
\lambda}\pi -3\dot{\lambda}H\Bigg),
\end{eqnarray}%
while the second takes the form
\begin{eqnarray}
3H^{2}+2{\dot{H}} &=&-\frac{p}{2\lambda }+\frac{1}{\lambda }\Bigg(-\frac{%
\kappa ^{2}f}{2}-\frac{1}{4}m^{2}\pi ^{2}-\frac{9}{8}{\lambda }\pi ^{2}-
\notag \\
&&\frac{9}{4}\dot{\lambda}\pi -2\dot{\lambda}H-\ddot{\lambda}\Bigg).
\end{eqnarray}

They are supplied with
\begin{equation}
\dot{\lambda}=\left( -\frac{2}{9}m^{2}-\kappa ^{2}f_{Q}-\lambda \right) \pi,  \label{lambdaevolution}
\end{equation}%
together with the constraint
\begin{equation}
\dot{\pi}=\frac{8}{3}H^{2}+\frac{4}{3}\dot{H}-3H\pi +\frac{1}{2}\pi ^{2}.
\end{equation}

By using \eqref{lambdaevolution}, we can eliminate the $\lambda $ terms from
the first and the second Friedmann equations. First, taking the derivative
of \eqref{lambdaevolution} gives
\begin{equation}
\begin{aligned} \ddot \lambda&=-\frac{2}{9} m^2 \dot \pi - \kappa^2 \dot
f_{Q} \pi - \kappa^2 f_{Q} \dot \pi - \dot \lambda \pi - \lambda \dot \pi
\end{aligned}
\end{equation}%
%
%
%
%
Hence, the second Friedmann equation can be rewritten as
\begin{eqnarray}
2\dot{H}+3H^{2}=-\frac{p}{2\lambda } &+&\frac{1}{\lambda }\Bigg(-\frac{%
\kappa ^{2}f}{2}-\frac{1}{4}m^{2}\pi ^{2}-\frac{9}{8}\lambda \pi ^{2}  \notag
\\
&&-\frac{9}{4}\dot{\lambda}\pi -2\dot{\lambda}H+\frac{2}{9}m^{2}\dot{\pi}%
+\kappa ^{2}\dot{f}_{Q}\pi   \notag \\
&&+\kappa ^{2}f_{Q}\dot{\pi}+\dot{\lambda}\pi +\lambda \dot{\pi}\Bigg).
\end{eqnarray}%
Further simplifying gives
\begin{eqnarray}
2\dot{H}+3H^{2} &=&-\frac{p}{2\lambda }+\frac{1}{\lambda }\Bigg(-\frac{%
\kappa ^{2}f}{2}-\frac{1}{4}m^{2}\pi ^{2}-\frac{9}{8}\lambda \pi ^{2}  \notag
\\
&&-\frac{5}{4}\dot{\lambda}\pi -2\dot{\lambda}H+\frac{2}{9}m^{2}\dot{\pi}%
+\kappa ^{2}\dot{f}_{Q}\pi   \notag \\
&&+\kappa ^{2}f_{Q}\dot{\pi}+\lambda \dot{\pi}\Bigg).
\end{eqnarray}%

\end{appendix}

\end{document}